\RequirePackage{rotating}
\documentclass[usenatbib]{mnras}
\usepackage[varg]{newtxmath}
\usepackage[T1]{fontenc}
\usepackage{ae,aecompl}
\usepackage{graphicx}
\usepackage{color}
\usepackage{journals}
\usepackage{comment}
\usepackage{multirow}
\usepackage{rotating}
\usepackage{xcolor,listings}
\usepackage{textcomp}
\usepackage{etoolbox}

\definecolor{grey}{rgb}{0.4,0.6,0.6}
\definecolor{brown}{rgb}{0.65,0.16,0.16}
\definecolor{darkgreen}{rgb}{0.0,0.6,0.0}

\begin{document}

\title [Formation and evolution of CGs]
{Compact groups from semi-analytical models of galaxy formation -- II:
   Different assembly channels}
\author[E. D\'iaz-Gim\'enez et al.]
{E. D\'iaz-Gim\'enez$^{1,2}$\thanks{eugenia.diaz@unc.edu.ar}, 
A. Zandivarez$^{1,2}$,
G. A. Mamon$^{3}$ 
\\
\\
$1$ Universidad Nacional de C\'ordoba (UNC). Observatorio Astron\'omico de C\'ordoba (OAC). C\'ordoba, Argentina\\
$2$ CONICET. Instituto de Astronom\'ia Te\'orica y Experimental (IATE). C\'ordoba, Argentina \\
$3$ Institut d'Astrophysique de Paris (UMR 7095: CNRS \& Sorbonne Universit\'e), Paris, France}

\date{\today}
\pagerange{\pageref{firstpage}--\pageref{lastpage}}
\maketitle
\label{firstpage}

\begin{abstract}
We study the formation of over 6000 compact groups (CGs) of galaxies  identified in mock redshift-space galaxy catalogues built from semi-analytical models of galaxy formation (SAMs) run on the Millennium Simulations. We select CGs of 4 members in our mock SDSS galaxy catalogues and, for each CG, we trace back in time the real-space positions of the most massive progenitors of their 4 galaxies. By analysing the evolution of the distance of the galaxy members to the centre of mass of the group, we identify 4 channels of CG formation. The  classification of these assembly channels is performed with an automatic recipe inferred from a preliminary visual inspection and based on the orbit of the galaxy with the fewest number of orbits.
Most CGs show late assembly, with the last galaxy arriving on its first or second passage, while only 10-20 per cent form by the gradual contraction of their orbits by dynamical friction, and only a few per cent forming early with little subsequent contraction. However, a SAM from a higher resolution simulation leads to earlier assembly.
Assembly histories of CGs also depend on cosmological parameters.
At similar resolution, CGs assemble later in 
SAMs built on parent cosmological simulations of high density parameter.  
Several observed properties of mock CGs  correlate with their assembly history: early-assembling CGs 
are smaller, with shorter crossing times, and greater magnitude gaps between their brightest two members, and their brightest galaxies have smaller spatial offsets and are more passive.
\end{abstract}

\begin{keywords}
galaxies: groups: general --
galaxies:  statistics --
methods: numerical
\end{keywords}

\section{Introduction}
\begin{table*}
\begin{center} 
\caption{Semi-analytical models analysed in this work}
\tabcolsep 5 pt
\begin{tabular}{lllcccclccccc}
\hline
\hline
\multicolumn{1}{c}{Authors\ \ } & \multicolumn{1}{c}{acronym} &\multicolumn{4}{c}{Cosmology} & & \multicolumn{2}{c}{Simulation} & &\multicolumn{3}{c}{Lightcone}\\ 
\cline{3-6} \cline{8-9} \cline{11-13}
& & \multicolumn{1}{c}{name} & $\Omega_{\rm m}$ & $h$ & $\sigma_8$ & & \multicolumn{1}{c}{name} & box size & & galaxies & CGs & CG4s\\
\multicolumn{1}{c}{(1)\ \ } & \multicolumn{1}{c}{(2)} &\multicolumn{1}{c}{(3)} & (4) & (5) & (6) & &\multicolumn{1}{c}{(7)} & (8) & & (9) & (10) & (11)\\
\hline
\cite{DeLucia+Blaizot07} & \ \ \ DLB & \ WMAP1 & 0.25 & 0.73 & 0.90 & & \ \ MS I & 500 & & 3\,757\,143 & 3387 & 1908 \\
\cite{Guo+11}       & \ \ \ G11 & \ WMAP1 & 0.25 & 0.73 & 0.90 & & \ \ MS I & 500 & & 3\,149\,024 & 3175 & 1570 \\
\cite{Guo+11}  		& \ \ \ GII & \ WMAP1 	   & 0.25 & 0.73 & 0.90 & & \ \ MS II & 100 & & 3\,214\,602 & 2558 & 1195 \\
\cite{Guo+13}       & \ \ \ G13 & \ WMAP7 & 0.27 & 0.70 & 0.81 & & \ \ MS I & 500 & & 2\,982\,462 & 1682 & 1010 \\
\cite{Henriques+15}$^{*}$  & \ \ \ H15 & \ Planck & 0.31 & 0.67 & 0.83 & & \ \ MS I & 480 & & 3\,087\,401 & 1291 & \ \,764 \\
\hline
\end{tabular}  
\end{center} 
\parbox{\hsize}{\noindent Notes:
The columns are:
(1): authors;
(2): acronym of SAM;
(3): cosmology of parent simulation;
(4): density parameter of parent simulation;
(5): dimensionless $z$=0 Hubble constant of parent simulation;
(6): standard deviation of the (linearly extrapolated to $z$=0) power spectrum on the scale of $8\,h^{-1}\,\rm Mpc$;
(7): name of parent simulation (MS I = Millennium Simulation I, MS II = Millennium Simulation II);
(8): periodic box size of parent simulation [$\, h^{-1} \, \rm Mpc$];
(9): number of galaxies in the mock lightcone;
(10): total number of compact groups identified in the mock lightcones ($r\le 17.77$);
(11): the number of compact groups with only four galaxy members.\\
More accurate values for the simulations can be found at
 \url{http://gavo.mpa-garching.mpg.de/Millennium/Help/simulation}.}
\parbox{\hsize}{\noindent  $^{*}$ H15 is run on a re-scaled
  version of the original Millennium simulation (hence, the different box size).}
\label{tab:sams}
\end{table*}

Isolated Compact Groups (CGs) of galaxies constitute an extreme environment, and a unique test-bed for galaxy interactions.
Indeed, these high apparent density and low velocity dispersions systems \citep{Hickson92} make them ideal laboratories for galaxy interactions and in particular mergers \citep{Mamon92}. In turn, these interactions induce  important transformations that have a strong impact on the history of their galaxy members.

For years, astronomers have been puzzled on how these small systems of galaxies can survive long periods of time and not ending up merging into a  single massive galaxy. Numerous $N$-body simulations have been carried out to  answer this question  (e.g. \citealt*{Carnevali+81,Barnes85,Mamon87,Navarro+87,Bode+93,Athanassoula+97}),  and generally agree that the galaxies rapidly merge over a timescale of 0.5 to 2 Gyr (see \citealt{Mamon90_summary} for a quantitative comparison of the results of the early simulations). If we observe CGs that do not survive long, a possible explanation for their existence today could be that the group membership is replenished by newly arriving galaxies  \citep*{Diaferio+94}, as expected from Press-Schechter \citep{Press&Schechter74} theory \citep{Mamon00_IAU174}. Nevertheless, under proper conditions, the merging process can be slowed down and therefore the longevity of these systems is not necessarily an issue. 

Still, it is not clear how such dense isolated galaxy systems form in the first place. If CGs seen in the local Universe were as dense when they formed as they appear now, they should have formed at early times ($z > 7$ according to Appendix~\ref{sec:zvir} or $z>3$ if the groups are double the size in real space as they appear in projection).   Alternatively, CGs may be galaxy systems that only very recently have acquired their fourth luminous member (CGs are defined to have at least 4 galaxies in a range of 3 magnitudes from the brightest group galaxy). In the same spirit, the fourth member may be on its second arrival. Finally, CGs may assemble when galaxies in a group lose their orbital energy and angular momentum by dynamical friction and spiral in to the inner part of virialized groups (e.g., \citealt{Schneider&Gunn82,Ponman+96}).

The assembly history of CGs is best explored with a very comprehensive numerical study. 
A suitable tool for such a study are the mock galaxy catalogues constructed using a semi-analytical model of galaxy formation (SAM) on top of a $N$-body cosmological simulation. 
During the last decade, such mock catalogues built from SAMs have been widely used to understand the frequency, nature and identification of CGs (\citealt*{McConnachie+08}; \citealt{DiazGimenez&Mamon10,Zandivarez+14,DiazGimenez+15,Taverna+16}; \citealt*{DiazGimenez+18}), as well as their mass assembly \citep{Farhang+17} and their frequency versus time \citep{wiens+19}.
The nature of CGs has also been analysed from mock CGs  extracted from hydrodynamical simulations \citep{Hartsuiker&Ploeckinger20}.

\cite{tzanavaris+19} analysed the evolutionary history of a handful of  groups identified in 3D as small, dense and isolated systems inspired in the observational criteria of CGs. Despite their small number of groups (one quintet, one quartet and 8 triplets), they visualised a variety of possible evolutionary paths suggesting that CGs form via different assembly channels, where all galaxies end up lying within 0.5 Mpc from the most massive galaxy within the last 9 to 1 Gyr.

We have recently performed a comprehensive study of the frequency and nature of CGs identified in nine mock lightcones constructed from five different SAMs  (\citealt{paper1}, hereafter Paper~I).
We found that the frequency and nature of CGs depend strongly on the cosmological parameters of the parent dissipationless cosmological simulation on which the SAM had been run.  The space density of physically dense CGs increases as a strong power of the amplitude of the power spectrum of primordial density fluctuations on the comoving scale of $8\,h\,\rm Mpc$, $\sigma_8$. Furthermore, the fraction of CGs that are not physically dense systems of at least 4 galaxies of comparable luminosity, but instead caused by chance alignments of galaxies along the line of sight, increases with the cosmological density parameter, $\Omega_{\rm m}$. In currently favoured cosmologies (e.g. WMAP9, \citealt{Hinshaw+13} and Planck, \citealt{PlanckCollaboration+20_cosmopars}), only roughly 35 per cent of CGs are physically dense (in rough agreement with early estimates by \citealt{Mamon86,Mamon87}). But higher-resolution $N$-body cosmological simulations capture better the formation of physically dense CGs, which now represent roughly half of all CGs.
In Paper~I, we also found that intrinsic differences in the SAM recipes also lead to differences in the frequency and nature of CGs.

In the present article, we address the following questions. Which  is the principal formation channel of CGs: did they assemble early and keep their density, did they gradually increase in density, or did the last galaxy arrive just now on its first or 2nd passage? Is there a mix of CG assembly scenarios? How do cosmological parameters affect CG formation? If CG formation channels are mixed, are there links between specific channels and observational properties of CGs?

To answer these questions, we explore the formation channels of CGs by analysing the mock CG samples we built from the Millennium Simulations \citep{Springel+05,BoylanKolchin+09} in Paper~I.
Using the galaxy merger trees from the SAMs, we then follow back in time the most massive progenitors of each of the member galaxies to study the spatial evolution of CGs. 

The layout of this work is as follows. In Sect.~\ref{sec:simulations}, we present the SAMs and the lightcone construction from these SAMs. We explain in Sect.~\ref{sec:sample} how we built the mock CG sample. Evolutionary tracks are illustrated and classified in Sect.~\ref{sec:histories}. The statistics of the formation scenarios are presented in Sect.~\ref{sec:stats}, while  their links to observational properties are explored in Sect.~\ref{sec:typeprops}. We conclude and discuss our
results in Sect.~\ref{sec:discus}.

\section{Mock Galaxy Lightcones}
\label{sec:simulations}

We used four SAMs run on the Millennium simulation (MS I, \citealt{Springel+05}): those of \cite{DeLucia+Blaizot07} and \cite{Guo+11} based on cosmologial simulations in the WMAP1 cosmology \citep{Spergel+03}; that of \cite{Guo+13} in the WMAP7 cosmology \citep{Komatsu+11}; and the SAM of \cite{Henriques+15} run on the Millennium simulation re-scaled to the Planck cosmology \citep{Planck+16}. These SAMs allow us to study the effect of cosmology on CGs formation channels.
Moreover, we also used the SAM of \cite{Guo+11} with the WMAP1 cosmology run on the Millennium Simulation II (MS II, \citealt{BoylanKolchin+09}) to understand the effects of space and mass resolution on the CG formation channels.
Table~\ref{tab:sams} lists the different SAMs with their cosmological parameters and parent simulations.

Mock CGs extracted from these SAMs were previously analysed to infer how CG properties depend on the underlying cosmological parameters, spatial and mass resolutions, as well as on the physical properties of the SAM (Paper~I).
Adopting the procedure of previous works  \citep{Guo+11,knebe15,irodotou19}, we only considered galaxies with stellar masses larger than $\sim 10^9 \, {\cal M}_\odot$ ($7\times 10^8 \, h^{-1} {\cal M}_\odot$) for the MS I SAMs, and stellar masses larger than $\sim 10^8 \, {\cal M}_\odot$ ($7\times 10^7 \, h^{-1} {\cal M}_\odot$) for the MS II SAM (see queries to retrieve these data in the Appendix of Paper~I). 

For each SAM listed in Table~\ref{tab:sams}, we constructed all-sky mock galaxy lightcones following a similar procedure as \cite{jpas}, using synthetic galaxies extracted from different outputs of the simulations to include the evolution of structures and galaxy properties with time. To compute the observer-frame apparent magnitudes we applied a k-decorrection procedure following the recipes described in \cite{DiazGimenez+18}. Our lightcones are limited to an apparent observer-frame Sloan Digital Sky Survey (SDSS) AB magnitude of $r \leq 17.77$. The total number of galaxies in the lightcone for each SAM is quoted in Table~\ref{tab:sams}. 
The stellar mass resolution of each SAM causes the luminosity distributions to be incomplete below a certain value (see Fig.~2 of Paper~I). Nevertheless, given the apparent magnitude limit used for building lightcones, the luminosity incompleteness is only present for a very small volume in the nearby universe ($\sim 0.35\%$ of the lightcone volume for the MS~I based SAMs).

\section{Compact group sample}
\label{sec:sample}

We use the samples of mock CGs that we previously identified in these mock galaxy catalogues (Paper~1). CGs were identified using the recent CG finding algorithm of \cite{DiazGimenez+18}. The CG finder looks for groups that simultaneously satisfy constraints of membership, flux limit of the brightest galaxy, compactness, isolation and velocity concordance. 
This algorithm is much more efficient, because previous algorithms, based upon selection of CGs projected on the sky \citep{Hickson82} followed by the removal of discordant redshifts \citep{Hickson92}, discard many CGs that appear non-isolated by the presence of neighbouring galaxies that would later turn out to have discordant redshifts. 
The algorithm also takes into account that the synthetic galaxies are point sized particles, whereas in galaxies lying very close on the sky in observed CGs may be blended (see \citealt{DiazGimenez&Mamon10}).  

We, hereafter, limit our analyses to CGs of 4 galaxies (hereafter, CG4s), because assembly histories are simpler when the final number of galaxies is the same.
Table~\ref{tab:sams} indicates that CGs of four galaxies constitute 47\% to 60\% of the total sample of CGs 
(decreasing with improved resolution).\footnote{In our samples of CGs, 
we deal with two different numbers of member galaxies: the \emph{observed} number and the \emph{real} number. 
The observed number of members takes into account the blending of galaxies that occurs when two particles in the simulation are too close in projection. 
For example, a CG with five members may appear as a CG of four observed members if one galaxy is blended with another. 
In this work, we use the real number of members when defining CG4s. }
The relative differences in the number of CG4s in each SAM resembles those of the CGs (including higher richness groups) previously analysed in Paper~I.

The Hickson-like automatic search identifies systems that can hardly be considered as dense structures in 3D space \citep{McConnachie+08,DiazGimenez&Mamon10}. 
These ``Fake'' groups have a large maximum 3D separation between their galaxies at $z=0$ ($>1\,h^{-1} \rm Mpc$) or they have no close pairs in 3D space (within $200\,h^{-1}\,\rm kpc$), which makes them unphysical systems.
Therefore, Fake CG4s will not be followed back in time to trace their history, and we will consider them as a separate class when needed.

\begin{figure}
\begin{center}
\includegraphics[width=0.95\hsize]{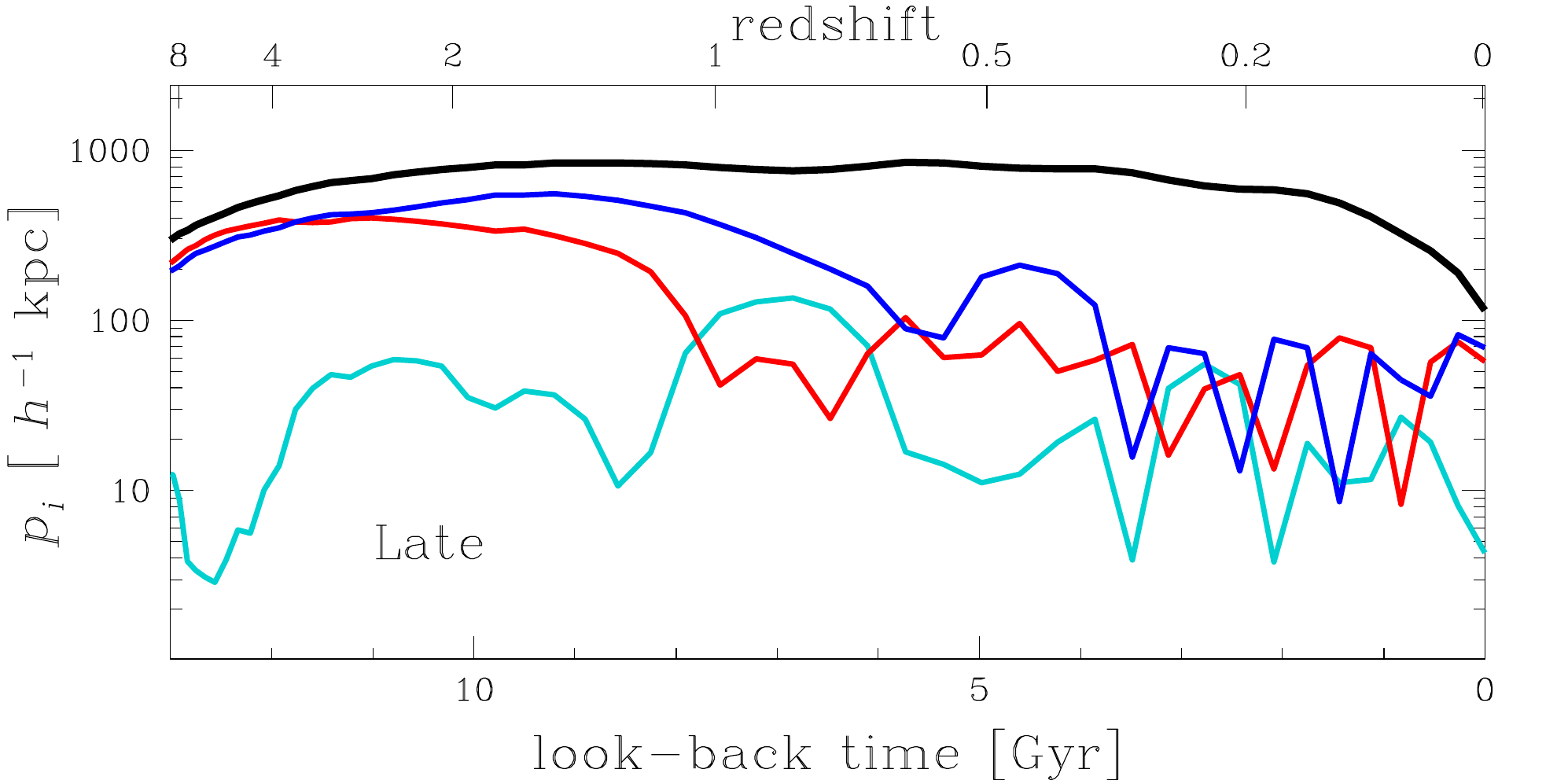}
\includegraphics[width=0.95\hsize]{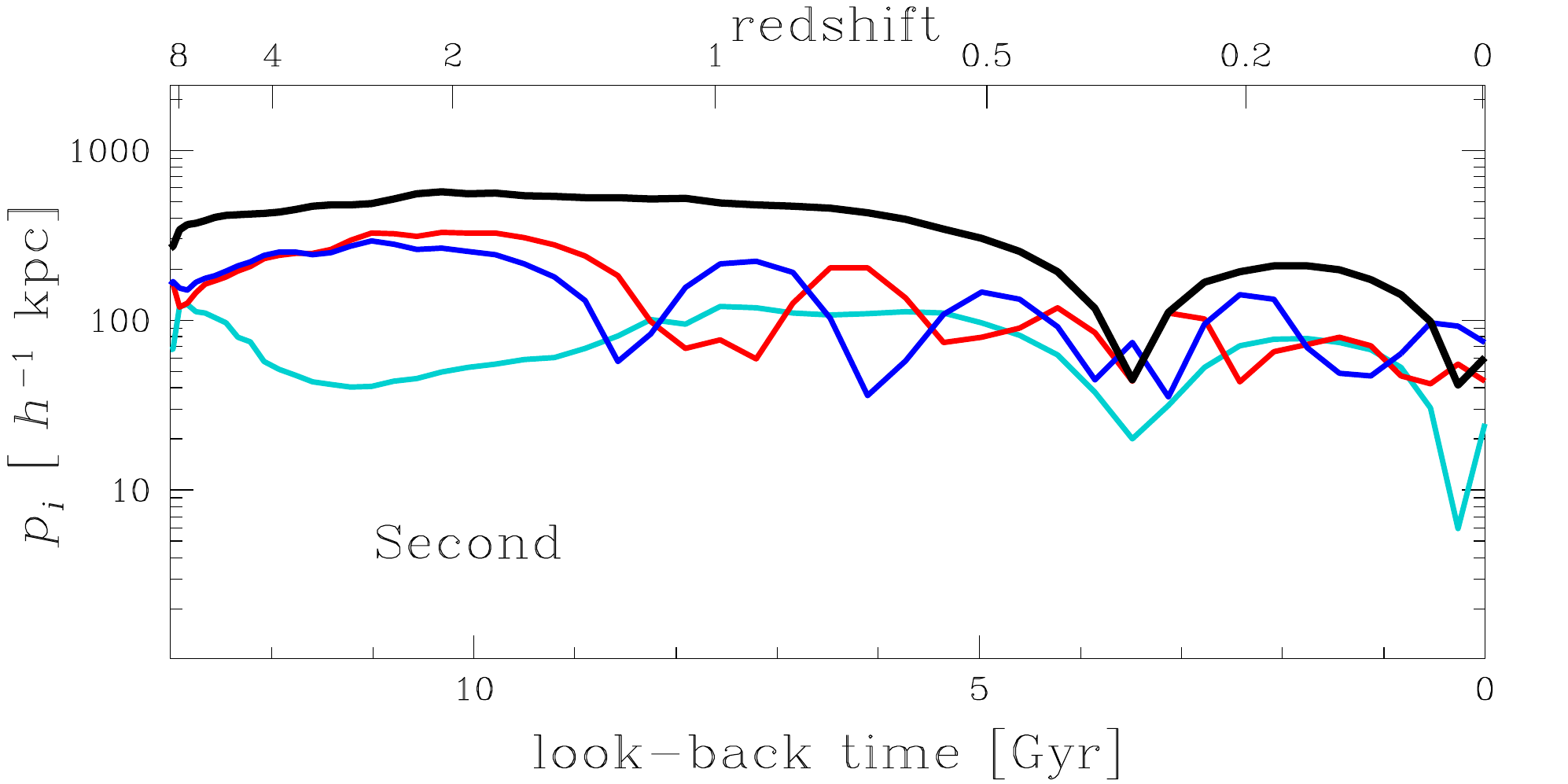}
\includegraphics[width=0.95\hsize]{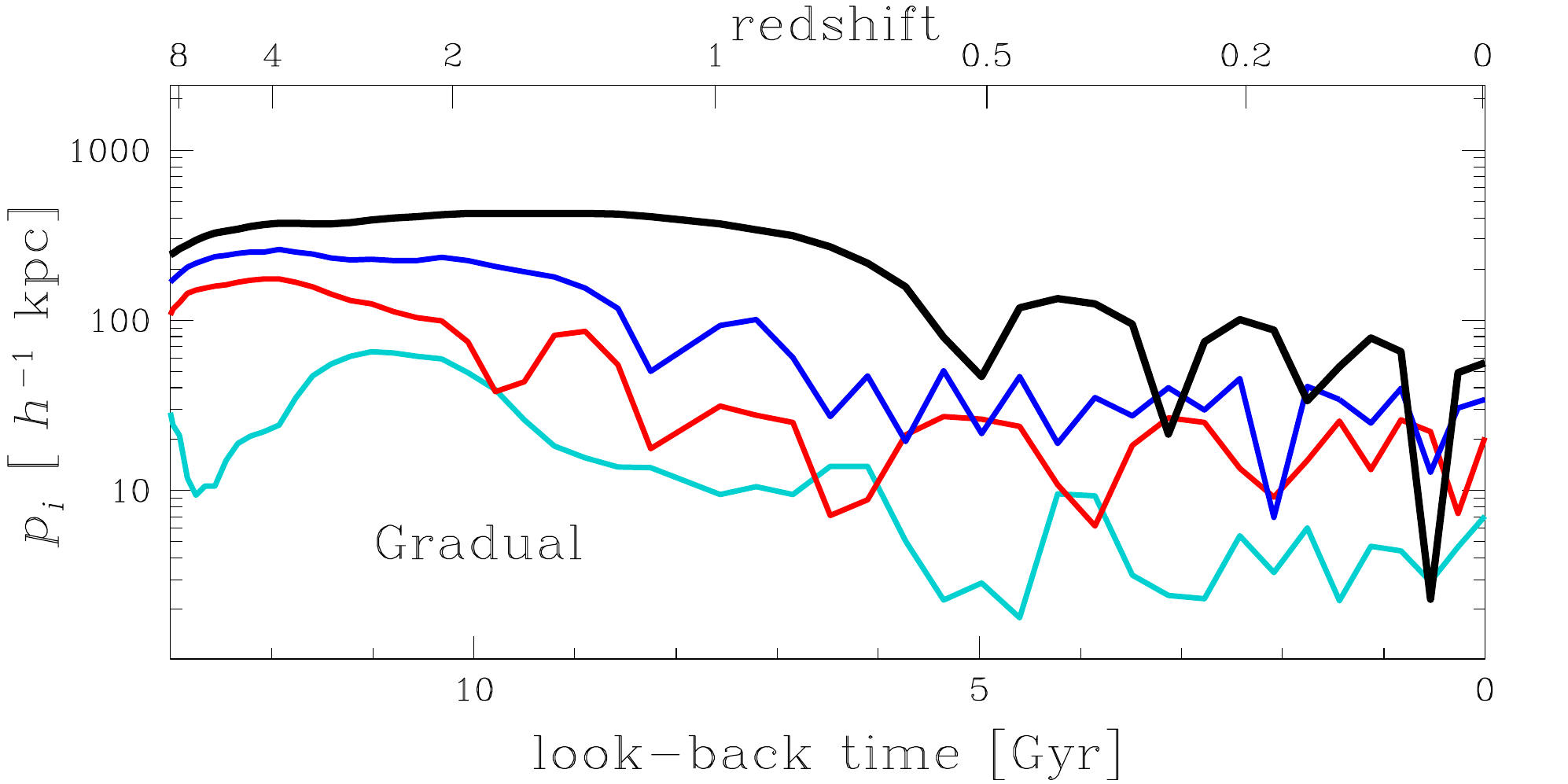}
\includegraphics[width=0.95\hsize]{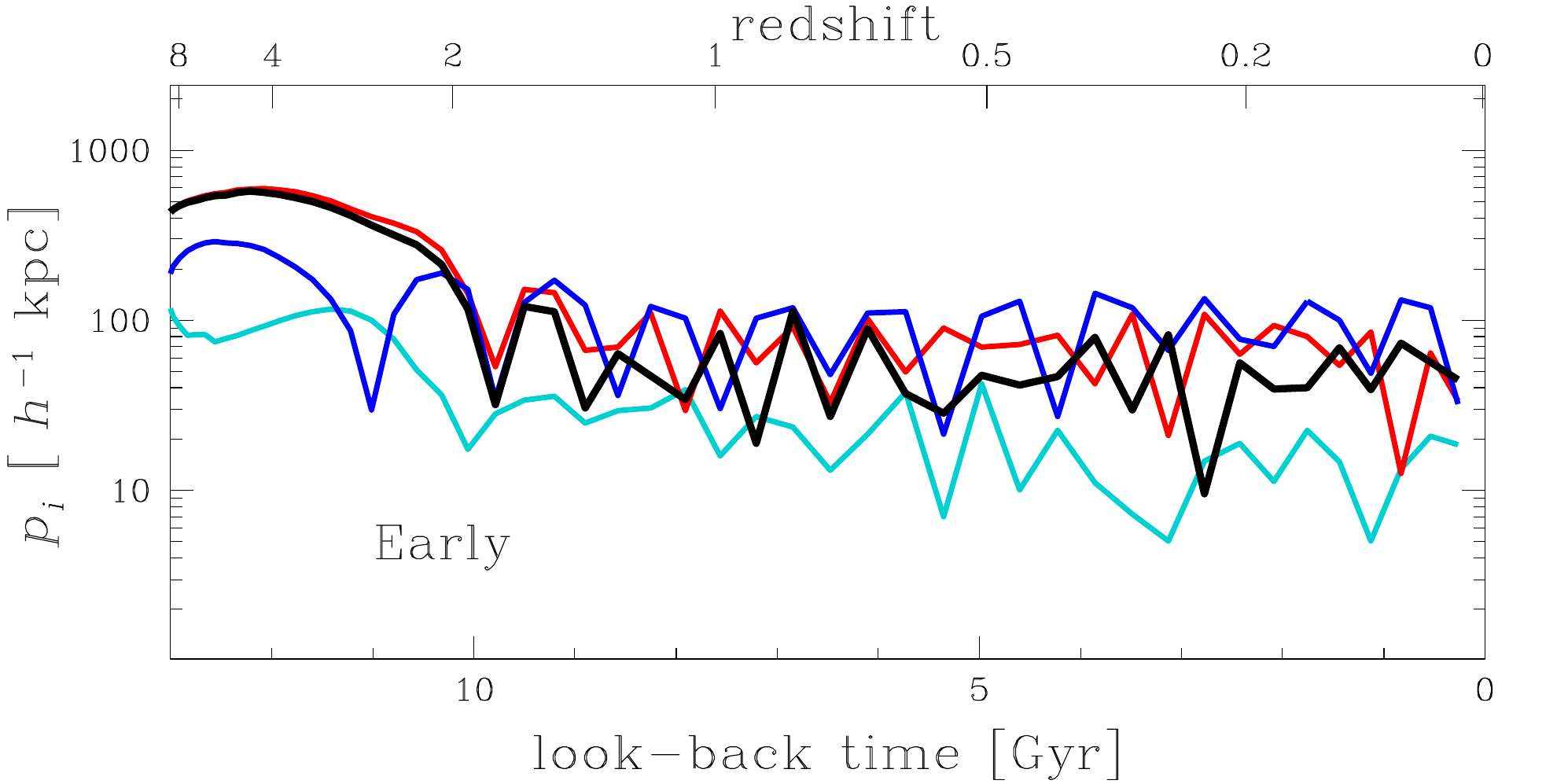}
\caption{
Assembly histories of four CG4s, defined by the main progenitors of their $z$=0 galaxies, traced by the evolution of the 3D physical distance of each galaxy member to the CG centre of mass. Time (age of the Universe) increases towards the right. The four rows highlight examples of the {\tt Late}, {\tt Second}, {\tt Gradual} and {\tt Early} CG assembly classes, from \emph{top} to \emph{bottom}. 
\emph{Black lines} display the key galaxies indicating the formation channel. 
The physical separation between two galaxies at a given look-back time is at most the sum of their $p_i$ values (since they may lie on opposite sides of the CG centre of mass). 
\label{fig:evoltest}}
\end{center}
\end{figure}

\section{CG assembly: a test sample}
\label{sec:histories}

Before analysing all 6447 CG4s, we 
restricted our analysis to a \emph{test sample} of 342 CG4s, built by selecting $\sim 5\%$ of the CG4s of each SAM. In the test sample, we excluded the Fake CGs. 

In our analyses, we tracked the evolutionary path of each $z$=0 galaxy member of the test sample. From the MS database\footnote{\url{http://gavo.mpa-garching.mpg.de/Millennium/}}, we retrieved the merger history of each of these galaxies using the query example G2 provided in the database. 
We then selected only the main progenitor of each $z$=0 galaxy (main branch of the merger tree) up to the redshift where the galaxy is formed. 

At each snapshot, CG4s are characterised by: 
\begin{itemize}
    \item the centre of stellar mass $ {\bf r}_{\rm cm} = \displaystyle  \frac{\sum_{i=1}^{n=4} {\cal M}^*_i\, {\bf r}_i }{\sum_{i=1}^{n=4} {\cal M}^*_i}$, where ${\bf r}_i$ is the Cartesian vector from the origin of the box to each galaxy member;
    \item the module of the comoving 3D distance of each member to the centre of mass. We name this parameter $d_{i}$; 
    \item the module of the physical 3D distance to the centre of mass computed as $\displaystyle p_{i}= \frac{d_{i}}{1+z_{\rm snapshot}}$; 
\end{itemize}

\subsection{Visual analysis of CG assembly histories}
\label{visual}

We observed four different types of CG assembly channels:
\begin{description}
\itemindent=0pt
\item{\bf {\tt Late Assembly}}: the fourth galaxy has just arrived in the CG for the first time (\texttt{Late}).
\item{\bf {\tt Late Second Pericentre}}: 
the fourth galaxy has arrived in the CG on its second passage
(\texttt{Second}).
\item{\bf {\tt Gradual Contraction}}: the four galaxies might or might not be together since an early epoch, but they have already completed two or more orbits together, becoming gradually closer with each orbit (\texttt{Gradual}).
\item{\bf {\tt Early Assembly}}: the four galaxies are together since an early epoch  (hereafter \texttt{Early}).
\end{description}

Figure~\ref{fig:evoltest} displays examples of these four CG assembly classes extracted from the test sample. 
The first example, {\tt Late Assembly}, shows that the last galaxy (in black) arrived from a physical distance of roughly $1\,h^{-1}\,\rm Mpc$.
The second example, {\tt Late Second Pericentre}, shows a dense triplet (all galaxies but the black one) forming around 8 Gyr ago ($z=1.2$) and gradually becoming denser with time, while the last arrival entered the group 3.5 Gyr ago (from a physical distance of $500   \,h^{-1}\rm kpc$), bounced out to nearly $200\,h^{-1}\,\rm kpc$ and  fell back into the group at $\sim z=0$, lying $40\,h^{-1}\,\rm kpc$ from the centre.
The third example, {\tt Gradual Contraction}, shows a CG that forms between 6 and 5 Gyr ago, and the orbits gradually decay. 
The fourth example, {\tt Early Assembly}, shows that the CG is already in place 9 Gyr ago, keeping its size ($p_{\rm max} \simeq 100\,h^{-1}\,\rm kpc$) since that time. 

From the visual classification of the test sample CG4s, we found that $32\%$ are {\tt Late}, $33\%$ are {\tt Second}, $24\%$ {\tt Gradual}, and $11\%$ {\tt Early}. 

Figure~\ref{fig:test} shows the frequency of the four CG assembly classes among the test sample of CG4s (blue solid lines), separately for our five considered SAMs.
This first attempt of classification shows a tendency to a predominant percentage of recently formed CG4s in those SAMs run on the MS I. In the GII SAM run on the MS II, the fractions of CG4s classified as {\tt Early} and {\tt Gradual} are enhanced by a factor two, and the {\tt Gradual} class is dominant.
However, the fractions of assembly classes found with the visually-classified test sample should be taken with caution, because of its small number of CGs.

\subsection{Automatic classification of CG assembly histories}
\label{sec:auto}

Our sample of 6000 CG4s was too large to fully classify by visual means.
Instead, we performed an automatic classification of CG assembly histories.
We  first measured the pericentres and apocentres of each galaxy's orbit relative to the centre of mass of the group. 
We defined pericentres as those points in the profiles where the steep changes from negative to positive, but also we only considered ``deep'' pericentres, i.e., there has to be a decay of 20\% in the height of the profile when considering the two previous and posterior snapshots. i.e., there is a pericentre in the snapshot $j$ if
    \begin{equation}
    \displaystyle
 \frac{p_{i}(t_{j-2})-p_{i}(t_{j})}{p_{i}(t_j)}>0.2 
    \hbox { AND } 
       \frac{p_{i}(t_{j+2})-p_{i}(t_j)}{p_{i}(t_j)}>0.2 
    \ .
    \label{peri}
    \end{equation}
    The apocentres are the highest values of $p_i(t)$ in the profiles between two consecutive pericentres. 

The profiles of each galaxy member are characterized by the following ten parameters: 

\begin{enumerate}
\item $t_{\rm 1p}$: time of the first (earliest) pericentre;
\item $h_{\rm 1p}$: $p_{i}$ at the time of the first pericentre;
\item $t_{\rm lp}$: time of the last pericentre;
\item $h_{\rm lp}$: $p_{i}$ at the time of the last pericentre;
\item $t_{\rm 1a}$: time of the first apocentre;
\item $h_{\rm 1a}$: $p_{i}$ at the time of the first apocentre;
\item $t_{\rm la}$: time of the last apocentre;
\item $h_{\rm la}$: $p_{i}$ at the time of the last apocentre;
\item $n_{\rm p}$: total number of deep pericentres (defined in eq.~[\ref{peri}]);
\item $p(t_0)/p(t_1)$: ratio between the $p_i$ in the two latest snapshots.
\end{enumerate}

Following the insights from the observations of the individual galaxy profiles, 
we characterize the assembly history of a CG4 by selecting the \emph{key galaxy} as the
 member with the fewest number of orbits, i.e. the lowest value of $n_{\rm p}$ (black lines in Fig.~\ref{fig:evoltest}). In those cases where two or more galaxies present the same number of pericentres, we defined the key galaxy as the galaxy with the most recent $t_{\rm 1p}$  provided that those times differ in more than 1 Gyr. When the latest requirement was not fulfilled, we defined the key galaxy as that with the highest $h_{\rm 1p}$. 
 
\begin{figure}
    \begin{center}
    \includegraphics[width=\hsize,viewport=0 30 550 550]{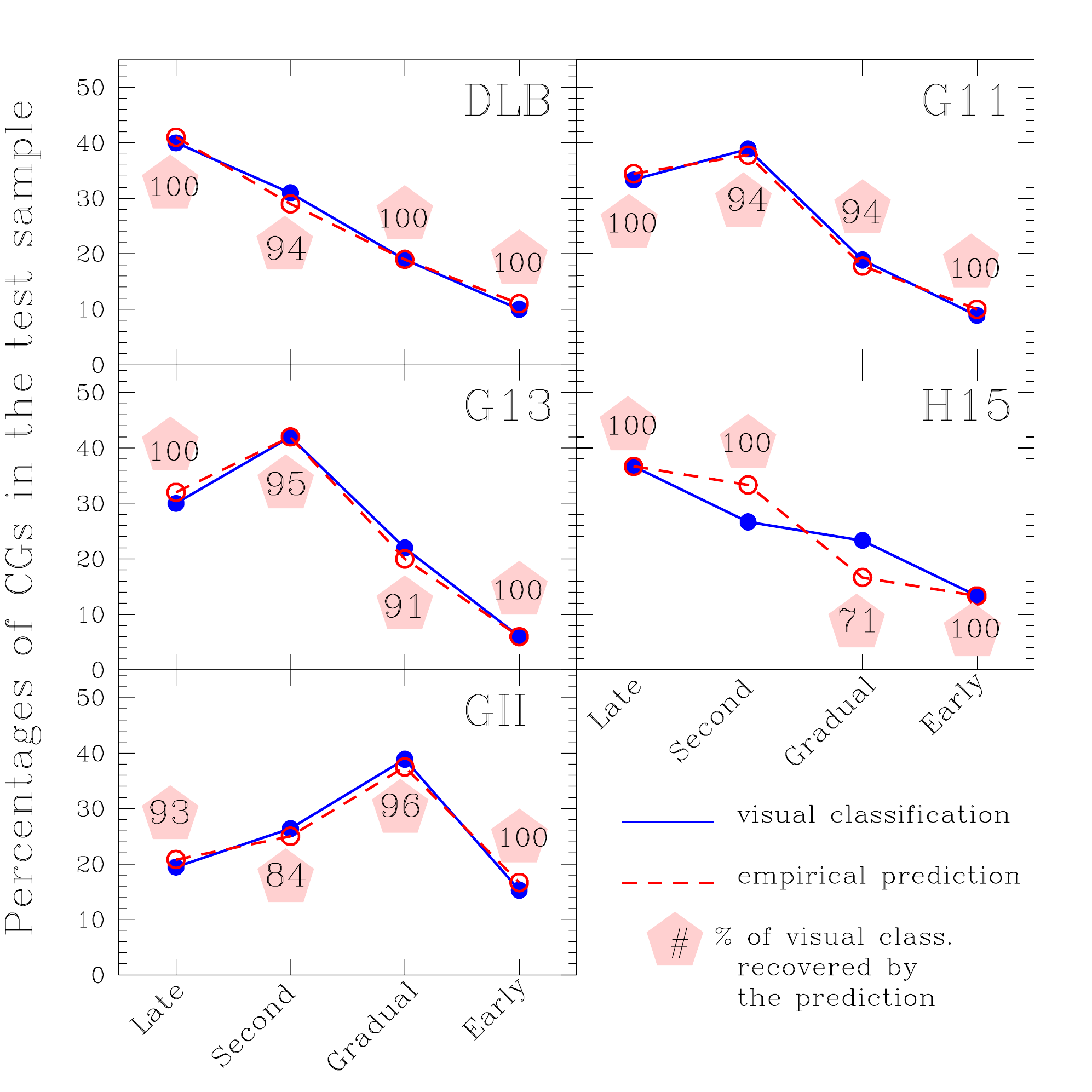}
    \caption{\label{fig:test}
    Distributions of assembly classes for the test sample of $z$=0 4-galaxy CGs, for the five different semi-analytical models.
    The results of the visual and empirical classifications are shown as \emph{blue filled circles and solid lines} and \emph{red open circles and dashed lines}, respectively.
    \emph{Inset numbers} show the percentages of the visual classification recovered by the automatic prediction in each class.
    }
    \end{center}
\end{figure}

\begin{figure}
    \begin{center}
    \includegraphics[width=\hsize,trim=0 30 0 0]{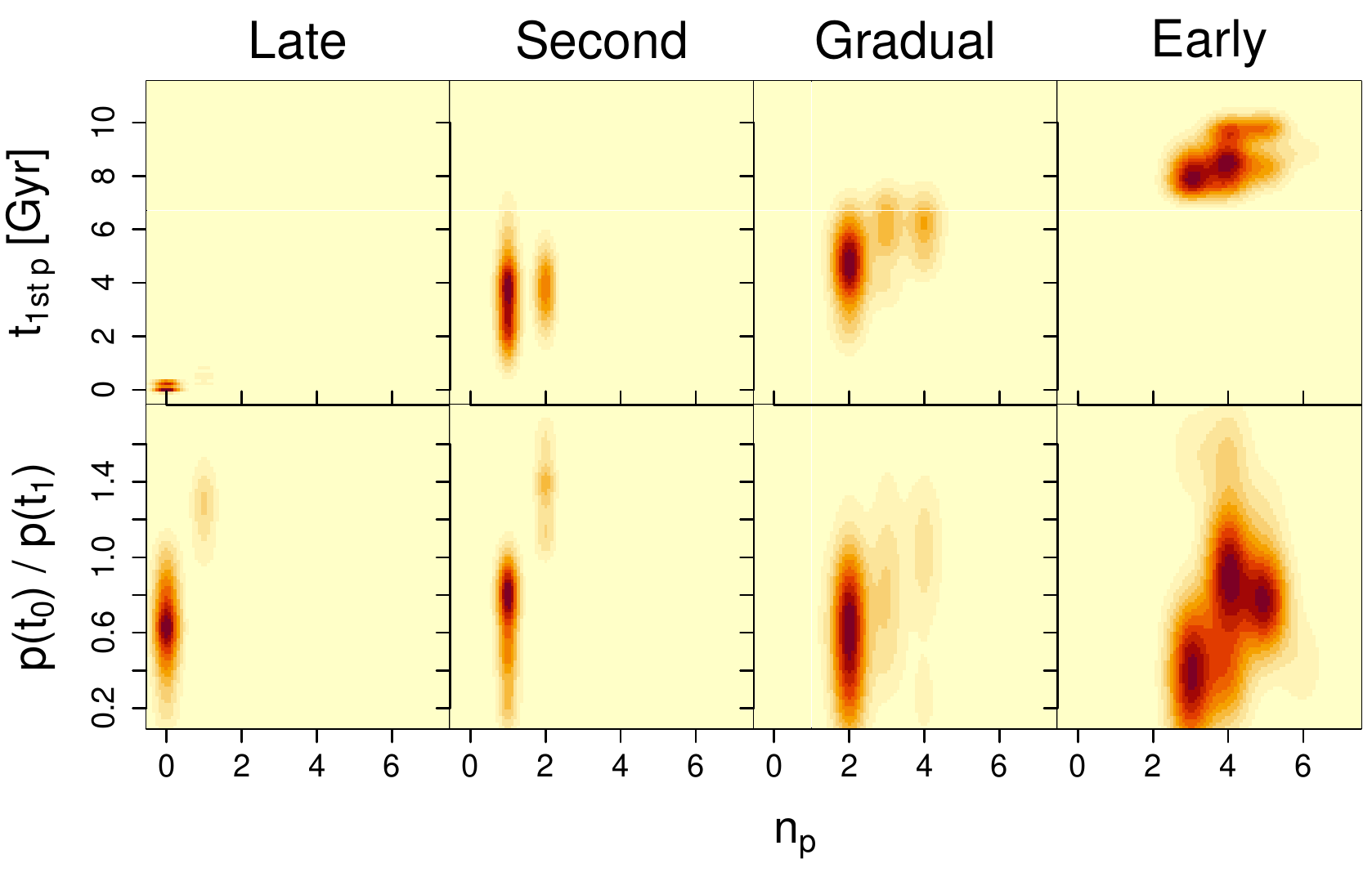}
    \caption{\label{fig:visual} 
    Heat diagrams for the scatter-plots of look-back time of the first pericentre (\emph{top row}) and ratio of the sizes of the profile at the latest times (\emph{bottom row}) both versus the number of pericentres, for the key galaxies of the test sample split into different assembly channels according to the visual classification.
    }
    \end{center}
\end{figure} 

Analysing different combinations of features in the profile of the key galaxy, 
we realize that our visual classification relies mainly on a few parameters of the profiles. 
In Fig.~\ref{fig:visual}, we show the scatter plots (as heat diagrams) of the number of pericentres in the key galaxy profile versus the look-back time of its first pericentre (top row), and versus the ratio $p(t_0)/p(t_1)$,
for groups split according to their visual classification (columns).
Based on these correlations, we propound the following automatic classification: 
\begin{description}
\itemindent 5pt
\itemsep 5pt
 \item[\tt Late:] $t_{\rm 1p}<7.5$ Gyr AND \\
 \phantom{t}\hfill \{$n_{\rm p}=0$ OR [$n_{\rm p}=1$ AND $p(t_0)>p(t_1)$]\} 
\item[\tt Second:] $t_{\rm 1p}<7.5$ Gyr AND \\ 
\{[$n_{\rm p}=1$ AND $p(t_0)\leq p(t_1)$] OR [$n_{\rm p}=2$ AND $p(t_0) > p(t_1)$]\} 
\item[\tt Gradual:]  $t_{\rm 1p}<7.5$ Gyr AND \\ 
\phantom{t}\hfill \{[$n_{\rm p}=2$ AND $p(t_0) \leq p(t_1)$] OR $n_{\rm p}>2$\} 
\item[\tt Early:] $ t_{\rm 1p} \ge 7.5$ Gyr
\end{description}
Note that this classification does not involve any physical size.

The application of this automatic classification to our test sample led to 96 per cent success at recovering the visual classification (the success rates within each SAM are 98, 97, 96, 93 and 93 for DLB, G11, G13, H15 and GII, respectively). 
The success rates as a function of visual class are 99, 93, 94 and 100 per cent for {\tt Late}, {\tt Second}, {\tt Gradual} and {\tt Early}, respectively.

Figure~\ref{fig:test} shows that the automatic classification applied to the test sample (red dashed lines) follows very well the visual classification (blue solid lines).
In Fig.~\ref{fig:test}, we also display
the percentage of visually-classified test CG4s that match the predictions from the automatic classification for each class of formation channels. 
These percentages of agreement are above 90 per cent and often 100 per cent, except for 2 among 20 cases: the {\tt Gradual} class with the H15 SAM (71 per cent) and the
{\tt Second} class with the GII SAM (84 per cent).
We, therefore, conclude that the predictions obtained from the automatic recipe reproduce the visual classifications of test CG4s with excellent fidelity.

\section{Results on compact group formation channels}
\label{sec:stats}
\begin{figure}
    \centering
    \includegraphics[width=\hsize,trim=0 0 20 30]{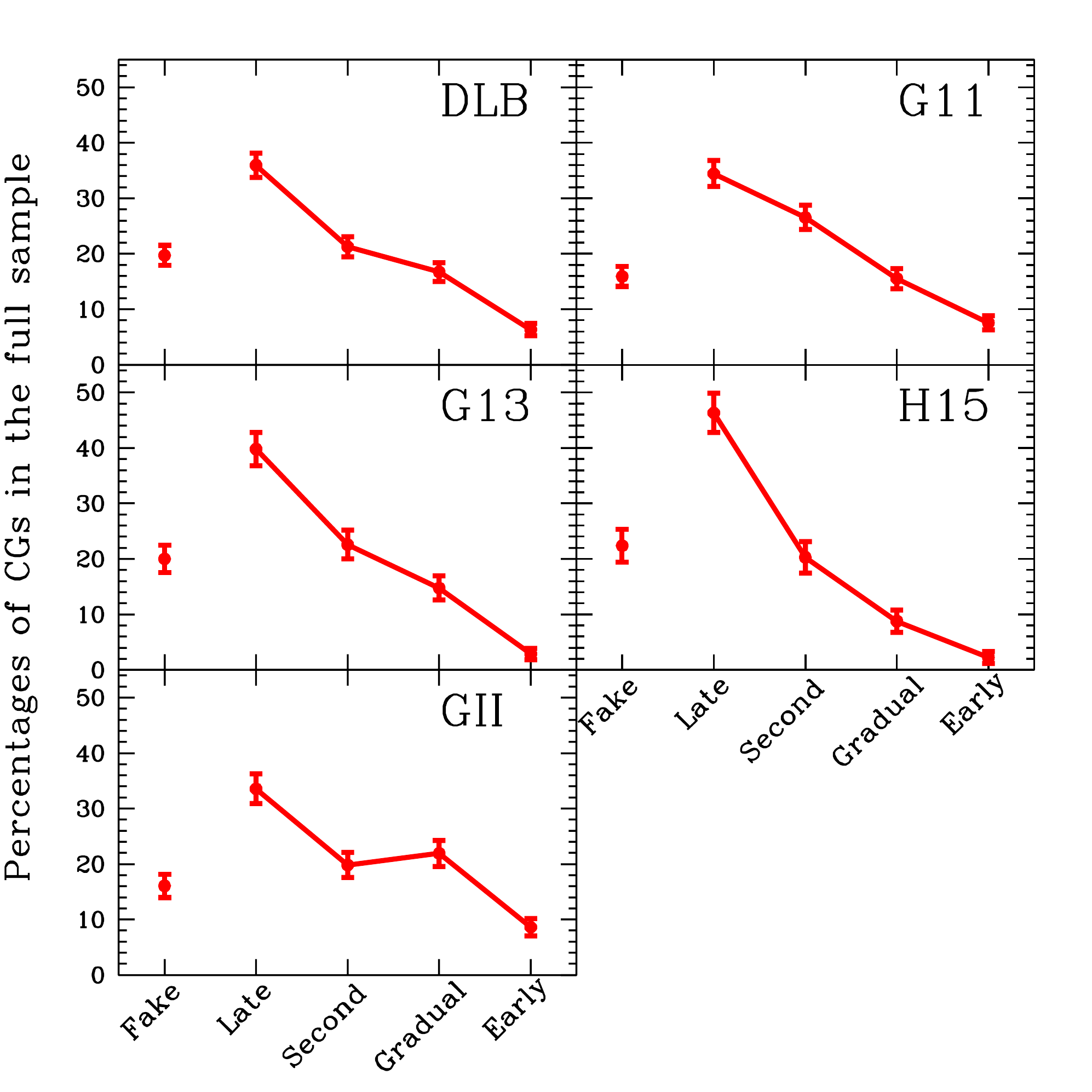}
    \caption{Percentages of $z$=0 4-galaxy CG assembly channels in the full sample, for the five different semi-analytical models, using the automatic classification.
    The percentages of CGs that are Fake are also shown. 
    Error bars are the 95\% binomial confidence intervals.
    }
    \label{fig:f4new}
\end{figure}

Given the high success rate achieved by the automatic classification, we applied it to the entire sample of CG4s.
Figure~\ref{fig:f4new} shows the percentage of CG4s within each formation channel. 
Roughly $20\%$ of the CG4s are classified as Fake. 
This figure shows some differences with those from the smaller test sample (dashed lines in Fig.~\ref{fig:test}), which suffers from worse statistics. However, regardless of the SAMs, the tendency of the predominant recent formation of CG4s is confirmed and the percentage of CGs diminishes towards the earlier assembly channels. Between one third and one half of the CG4s are {\tt Late}, while less than a tenth are {\tt Early} systems.

\subsection{Physically dense and chance alignments CGs}
We split the samples of CG4s into \emph{physically dense} \nobreak{(Reals)} and \emph{chance alignments} (CAs) following the classification of \cite{DiazGimenez&Mamon10}. 
Real CGs are those whose  maximum comoving 3D separation between the four closest galaxies is less than $100 \, h^{-1} \, \rm kpc$, or less than $200\,h^{-1}\,\rm kpc$ if the the ratio of line-of-sight to transverse sizes in real space is  less than $2$. The remaining CGs are classified as CAs. In this scenario, the Fake CGs defined in Sect.~\ref{sec:sample} belong to the CA subsample.

\begin{table}
\tabcolsep 2.5pt
\centering
\caption{Number of $z$=0 CGs with only four members for each class}
\begin{tabular}{lrrrrrrr}
    \hline
         SAM & \multicolumn{1}{c}{CG4} && \multicolumn{2}{c}{Reals} && \multicolumn{2}{c}{CAs}  \\
         \cline{4-5}
        \cline{7-8}
         & \multicolumn{1}{c}{Total} && \multicolumn{1}{c}{Total} & \% && \multicolumn{1}{c}{Total} & Fake \\
        \hline
All & 6447 && 2024 & 31 && 4423 & 1191 \\
\hline
DLB & 1908 && 588 & 31 && 1320 & 376 \\
G11 & 1570 && 530 & 34 && 1040 & 250 \\
G13 & 1010 && 300 & 30 && 710 & 202 \\
H15 & 764 && 185 & 24 && 579 & 171 \\
GII & 1195 && 421 & 35 && 774 & 192 \\
    \hline
    \hline
\end{tabular}
\label{tab:cas}
\end{table}

The numbers of different classes of CG4s identified in each lightcone are listed in Table~\ref{tab:cas}. 
One first notices that the fraction of Reals among CG4s is typically 30 per cent lower than the corresponding fraction among all CGs given in the top panel of fig.~8 of Paper~I. In particular, among CG4s, the fraction of Reals never exceeds 35 per cent (reached for GII).

\begin{figure}
    \centering
    \includegraphics[width=\hsize,trim=0 0 20 0]{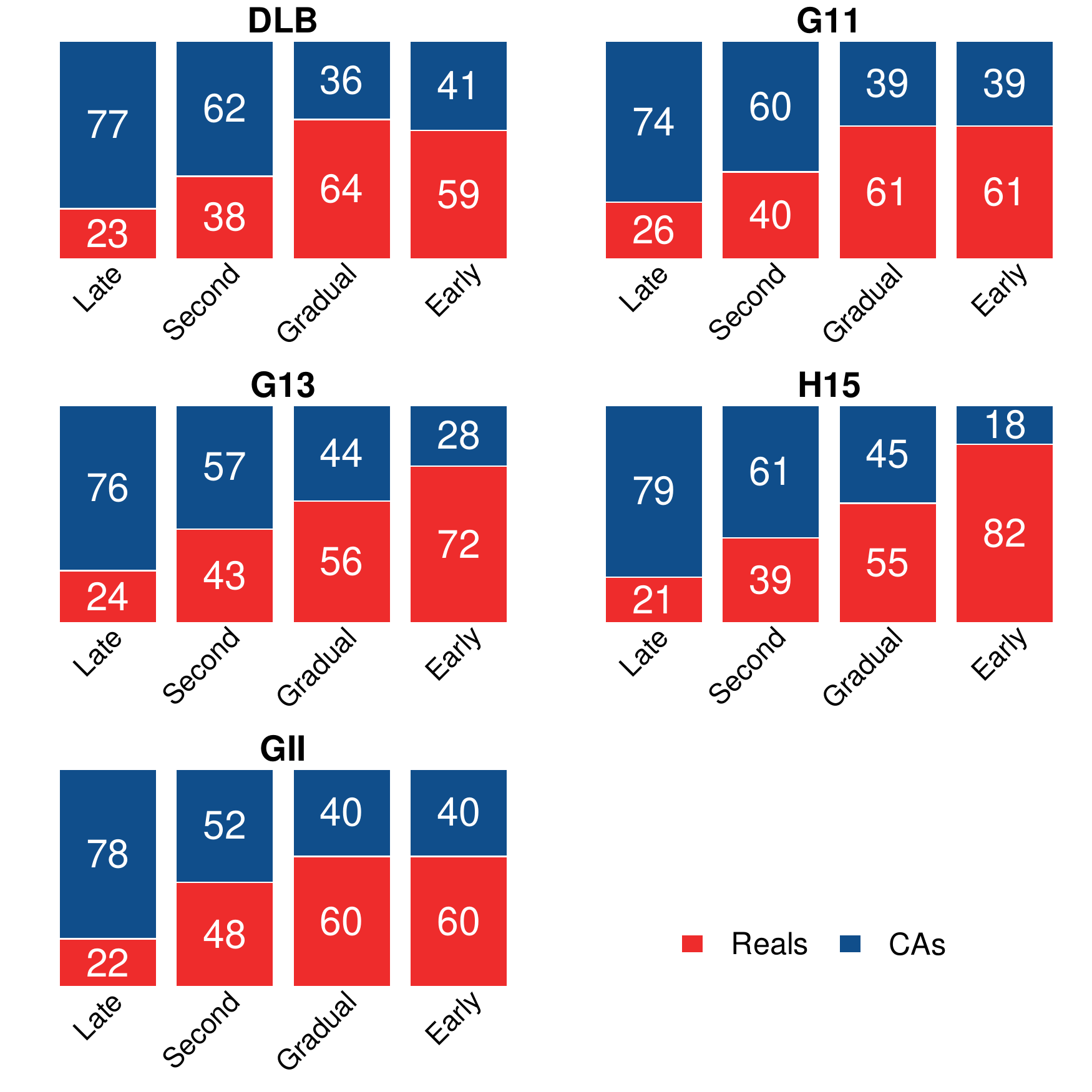}
    \caption{
        Percentage of Reals and CAs (including Fakes) within each formation channel. 
        }
    \label{fig:split}
\end{figure}
In addition, in Fig.~\ref{fig:split}, we show the percentage of Reals and CAs within each formation channel. Regardless of the SAM, over three-quarters of the {\tt Late} forming CGs are CAs; CAs are also the dominant population within the {\tt Second} channel, while the Reals dominate the {\tt Early} assembly CGs. 
In Appendix~\ref{nnfull}, we focused on the CA CG4s by analysing their internal substructures. 
Among the few CAs that formed earlier, the CAs formed by a triplet plus a somewhat distant galaxy are the main contributors to the long lived classes (Fig.~\ref{fig:cassub}).

\begin{figure}
    \centering
    \includegraphics[width=\hsize,trim=0 0 20 30]{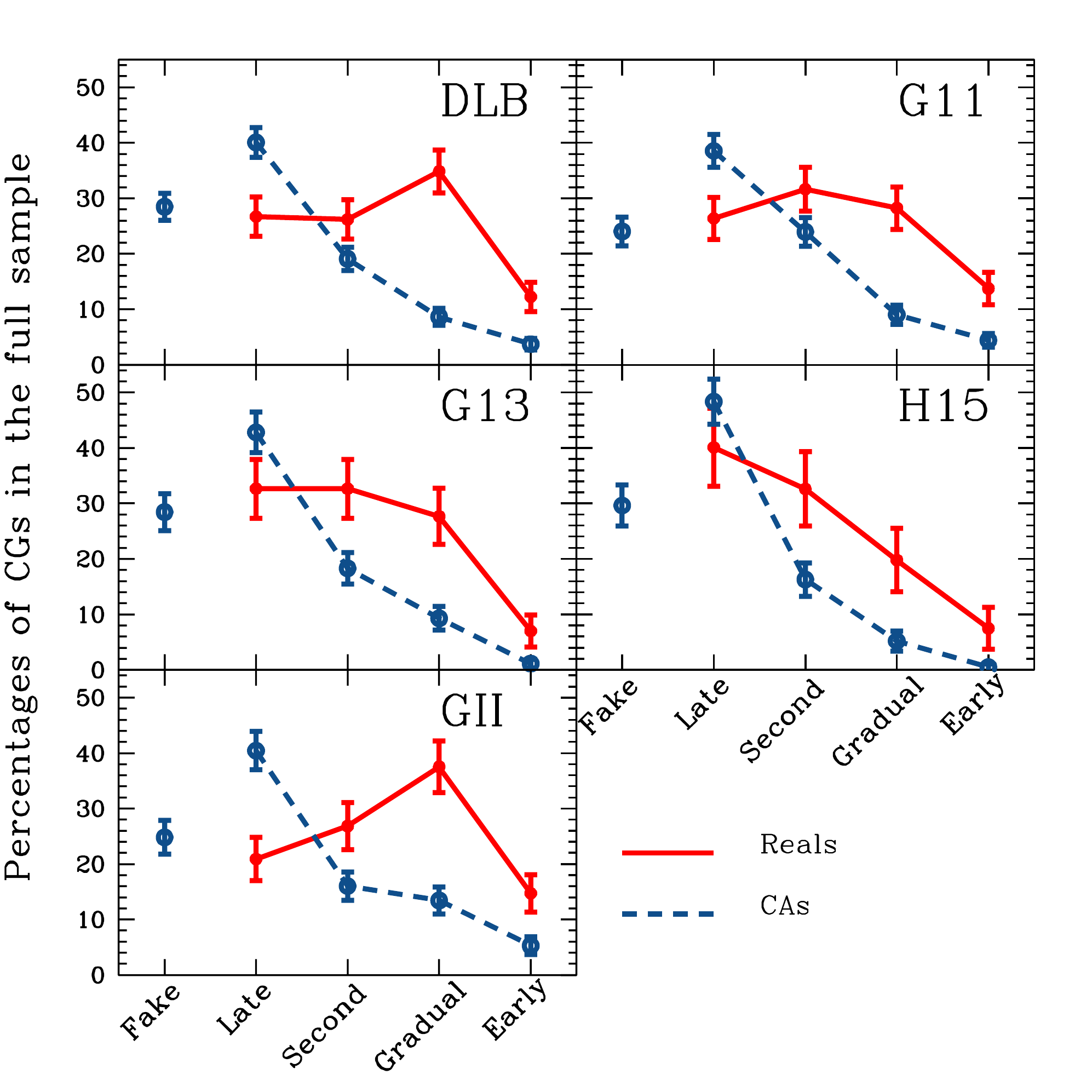}
    \caption{Percentages of $z$=0 4-galaxy CG assembly channels for Real and chance-alignment CGs (including Fakes) in the full sample (\emph{red} and \emph{blue}, respectively), for the five different semi-analytical models, using the automatic classification.
    The percentages of CAs that are Fake are also shown. 
    Error bars are the 95\% binomial confidence intervals.
    }
    \label{fig:full}
\end{figure}

Figure~\ref{fig:full} displays the percentages of
different CG assembly channels among Reals and CAs. 
The assembly histories of Real CGs depends somewhat on the SAM.
Real CGs in the DLB and GII SAMs are mainly assembled gradually, while 
in G11 and G13, Real CGs present a homogeneous mixture of {\tt Late}, {\tt Second} and {\tt Gradual} classes.
Finally, Real CGs in the H15 SAM are dominated by recently assembled CGs.

Figure~\ref{fig:full} indicates that, regardless of the SAM, CAs assemble significantly more recently than Real CG4s.
Indeed, the {\tt Late} class of assembly history accounts for a significantly higher fraction of CAs (typically over 40 per cent) than of Reals (typically under 30 per cent).

\subsection{Dependence on cosmology and SAM recipes}
\begin{figure*}
    \centering
    \includegraphics[width=0.7\hsize,viewport=10 25 554 554]{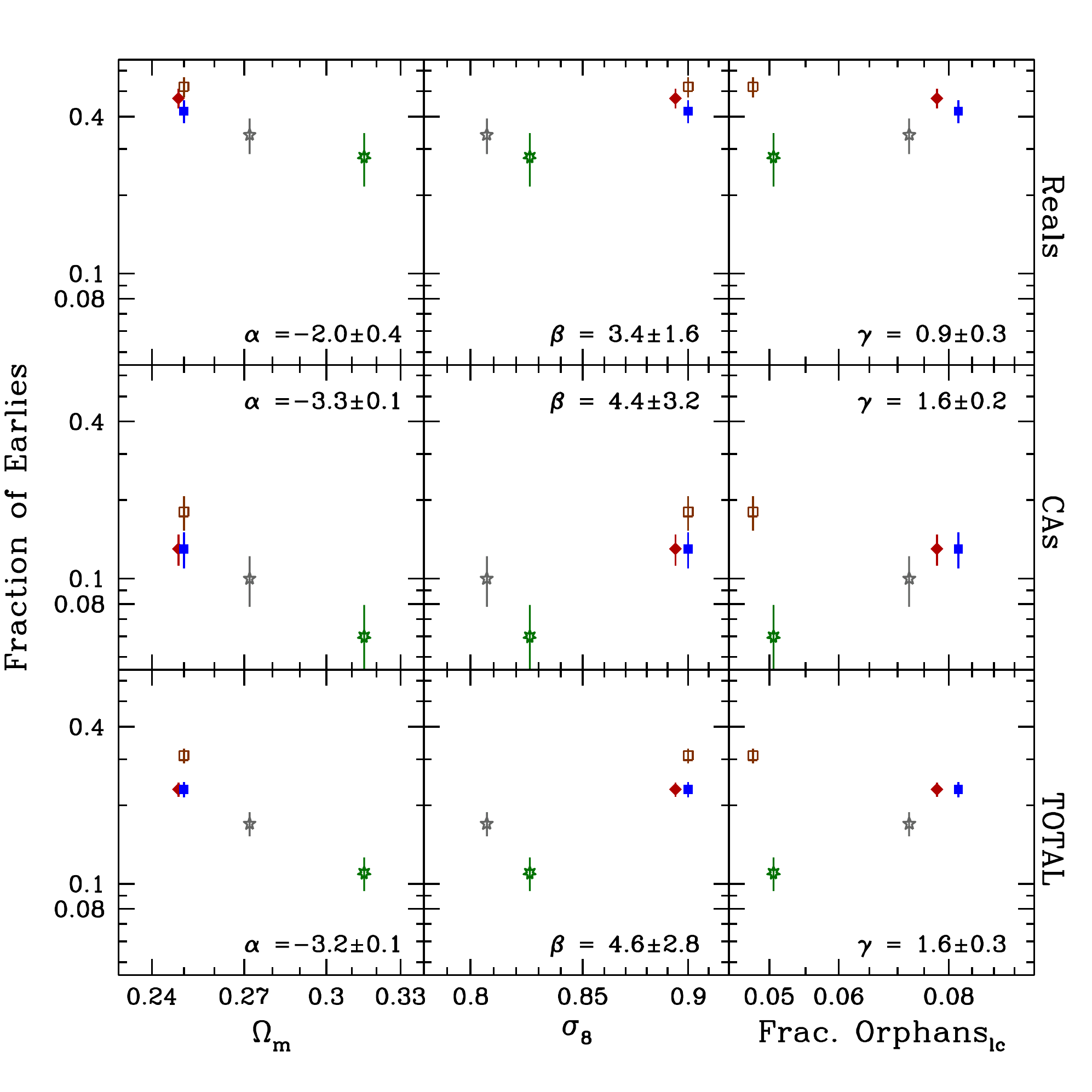}
    \includegraphics{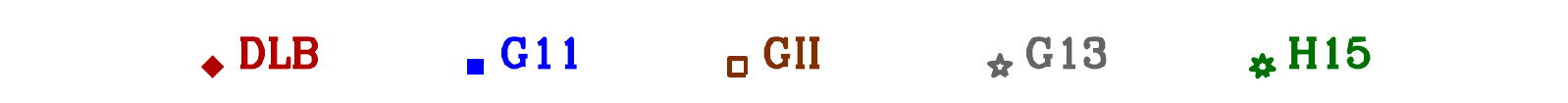}
    \caption{Fraction of $z$=0 4-galaxy CGs whose assembly histories are {\tt Earlies} ({\tt Gradual}+{\tt Early}) as a function of the cosmological matter density parameter
    ($\Omega_{\rm m}$) (\emph{left panels}), the amplitude of the mass fluctuations in a sphere of $8 \,h^{-1}\, \rm Mpc$ ($\sigma_8$) (\emph{middle panels}), and the fraction of orphan galaxies in the lightcones (\emph{right panels}). The trends are shown for the subsamples of Real and CA CGs (\emph{upper and middle panels}) as well as for the total sample (\emph{lower panels}). The normalizations of CA and total samples include the Fake CGs.
    In each panel, the values of the slopes of the fits using all MS~I are quoted. 
}
    \label{fig:earlies}
\end{figure*}

Following Paper~I, 
we explore the dependence of our results with the underlying cosmological parameters of the SAMs and with their physical recipes. Figure~\ref{fig:earlies} shows the fraction of a composite assembly class named \texttt{Earlies}, which combines {\tt Gradual} and {\tt Early} classes, as a function of cosmological parameters and the fraction of orphan galaxies in each SAM. 
We estimate the power-law relations for the fraction of {\tt Earlies}: $ \Omega_{\rm m}^\alpha$,  $ \sigma_8^\beta$, and $ f_{\rm orph}^\gamma$. The fits were obtained using all the SAMs with the same resolution (MS~I). The best-fitted values of the exponents are quoted in the figure. 

We first noted some trends of CG assembly classes with cosmological parameters. SAMs built in denser universes (G13 and especially H15) have lower fractions of {\tt Earlies}. 
On the other hand, there is only a weak positive trend of the assembly history as a function of the amplitude of mass fluctuations ($\sigma_8$). 
We discuss these trends in Sect.~\ref{sec:discus} below.

\cite{Contreras13} and \cite{pujol17}
observed that differences in the treatment of orphan galaxies in the SAMs impact on the small scale clustering of galaxies, which are the relevant scales for CGs.
Orphan galaxies are those whose subhaloes became no longer sufficiently massive to be realistic or no longer visible at all.
The SAMs can no longer use the subhaloes to follow their positions and rely instead on crude recipes for their orbital evolution. Hence, the treatment of orphans implemented in each SAM affects the abundance of this particular population of satellite galaxies. 

In Paper~I, we noticed that orphan galaxies are frequent in CGs, especially Real ones.
We also found that the fraction of CGs that are physically dense (i.e. Reals)  decreases with the fraction of orphans in the SAMs for given cosmological parameters, but increases with orphan fraction for given resolution.
We therefore analysed
the fraction of {\tt Earlies} as a function of the fraction of orphan galaxies in the lightcones built from each SAM (right panels of Fig.~\ref{fig:earlies}). 

Considering SAMs obtained from the MS I simulation (i.e., all with the similar mass resolution), Fig.~\ref{fig:earlies} shows that the fraction of {\tt Earlies} increases with the fraction of orphan galaxies. This trend appears to be statistically significant in Reals and even more pronounced in CAs.
As noted in Paper~I, lower fractions of orphan galaxies are found in mocks from SAMs run on the
 higher resolution MS~II simulation. Indeed, such simulations allow following the orbits of low-mass galaxies, which tend to live in lower subhalo masses, instead of having to predict them with uncertain dynamical recipes. 
From this figure, it can be seen that despite the lower fraction of orphans, a higher fraction of {\tt Earlies} is found in the MS II,
similar to the fraction of {\tt Earlies} in the comparable G11 SAM.
This suggests that orphans play a minor role in comparison with $\Omega_{\rm m}$ in setting the fraction of CG4s assembling early.

\begin{figure}
    \begin{center}
     \includegraphics[width=\hsize,trim=0 0 20 30]{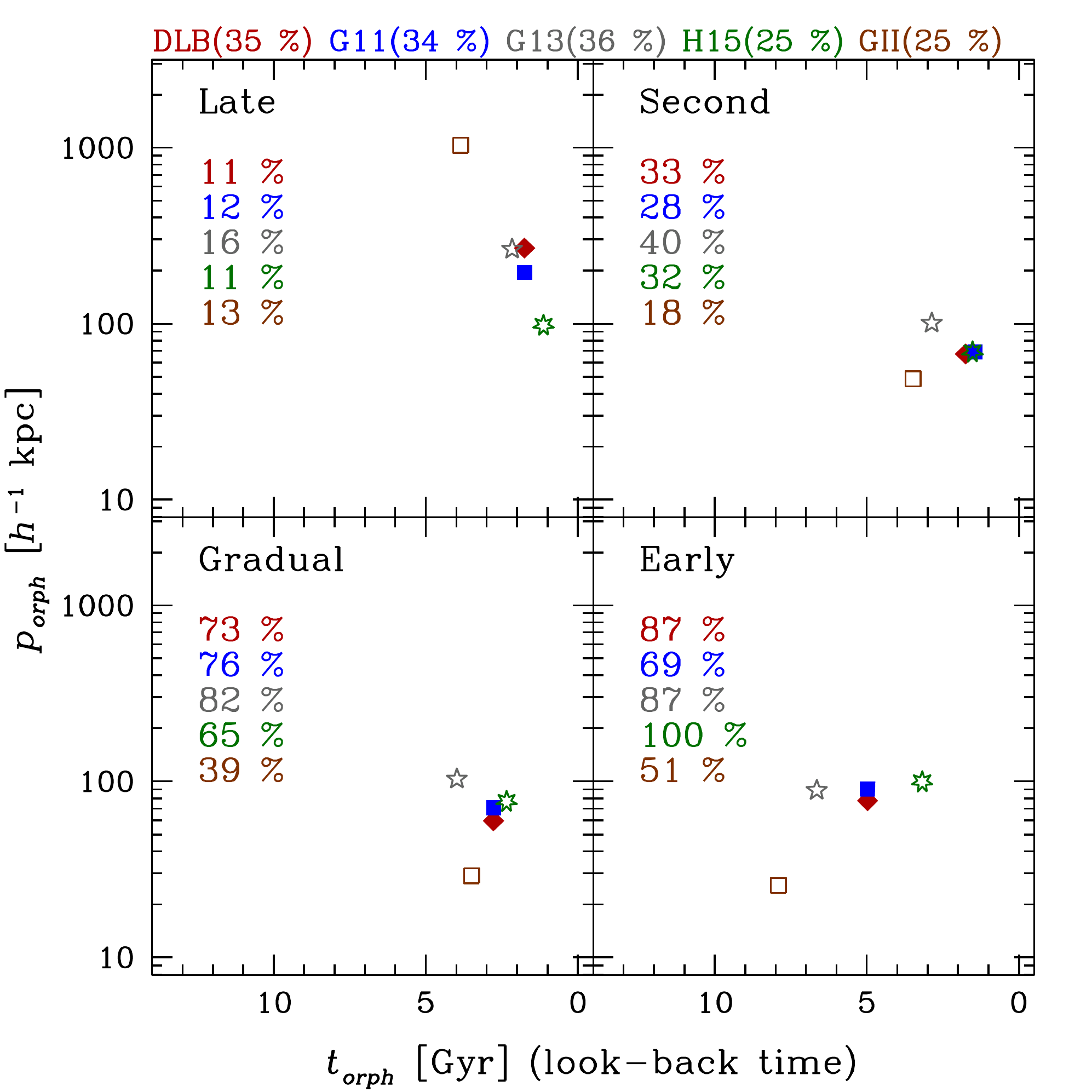}
    \caption{\label{fig:orph} 
    {Median (over CGs) positions and times when key galaxies become orphan (they lose their subhalos and the SAMs treat their orbits within analytical recipes).
    \emph{Top labels} quote the percentages of CG4s (excluding Fake CGs) in each SAM whose key galaxies became orphans. Each panel corresponds to the different assembly channels. Different \emph{symbols} display the different SAMs, as in Fig.\ref{fig:earlies}. \emph{Inset legends} quote the percentage of  CG4s within each assembly channel whose key galaxy became an orphan at some time (from top to bottom: DLB, G11, G13, H15 and GII).   
    }
    }
    \end{center}
\end{figure}

Our classification is based on the orbits of key galaxies,
some of which require analytical recipes when the subhalo has too few particles or disappears entirely, i.e. for orphan galaxies. We therefore examined the impact of orphan key galaxies within each formation channel, and we show the main results in Fig.~\ref{fig:orph}.

First, we found that, summing over all assembly channels, roughly one-third (or less depending on the SAM) of the key galaxies become orphans at some time in their evolutionary history (top labels). 
Furthermore, the occurrence of orphan key galaxies increases from the {\tt Late} to the {\tt Early} channel. 
For the MS~I based SAMs, the percentage of key galaxies that are orphans rises from $\sim 13\%$ in the {\tt Late} CGs to $\sim 85\%$ in the {\tt Early} CGs. For the better resolved GII SAM, this increase is considerably lower.

From the  merger trees of the orphan key galaxies, it is possible to determine the lookback-time and the physical distance to the mass centre when they became orphans. The median values of these two features are shown in Fig.~\ref{fig:orph}. 
Typically, the key galaxies in the {\tt Second} and {\tt Gradual} channels become orphans  during the last 5 Gyr, when they are less than $100   \,h^{-1}\,\rm kpc$ from the CG centre (typically less than $50\,h^{-1}\,\rm kpc$ for the better resolved GII SAM).
The key galaxies in the {\tt Early} channel have also become orphans when they are close to the centre, but this occurs earlier. In these three assembly classes,  galaxies become orphans at  distances to the CG centre-of-mass of order the half-maximum size of the system ($200\,h^{-1}\,\rm kpc$) or less (in particular for the better resolved GII SAM, but also for some other cases).
Therefore, the assembly into the small CG size typically occurs when the galaxies are not yet orphans, i.e. before the time when the SAMs use analytical recipes to predict galaxy orbits. 

On the other hand, the key galaxies in the {\tt Late} channel became orphans only recently, but at larger distances to the centre.
However, the percentage of {\tt Late} CGs that are classified using an orphan galaxy is very low in all the SAMs.

We conclude that the presence of orphan galaxies as key galaxies does not modify the results found in this work. 
 We also checked whether the visual classification was influenced with the key galaxy being an orphan,  but found no clear such influence.

\section{Are observed CG properties linked to their assembly history?} 
\label{sec:typeprops}
One may wonder whether observational properties of CG4s may retain signatures of their particular assembly history.

\begin{figure}
    \begin{center}
      \includegraphics[width=\hsize]{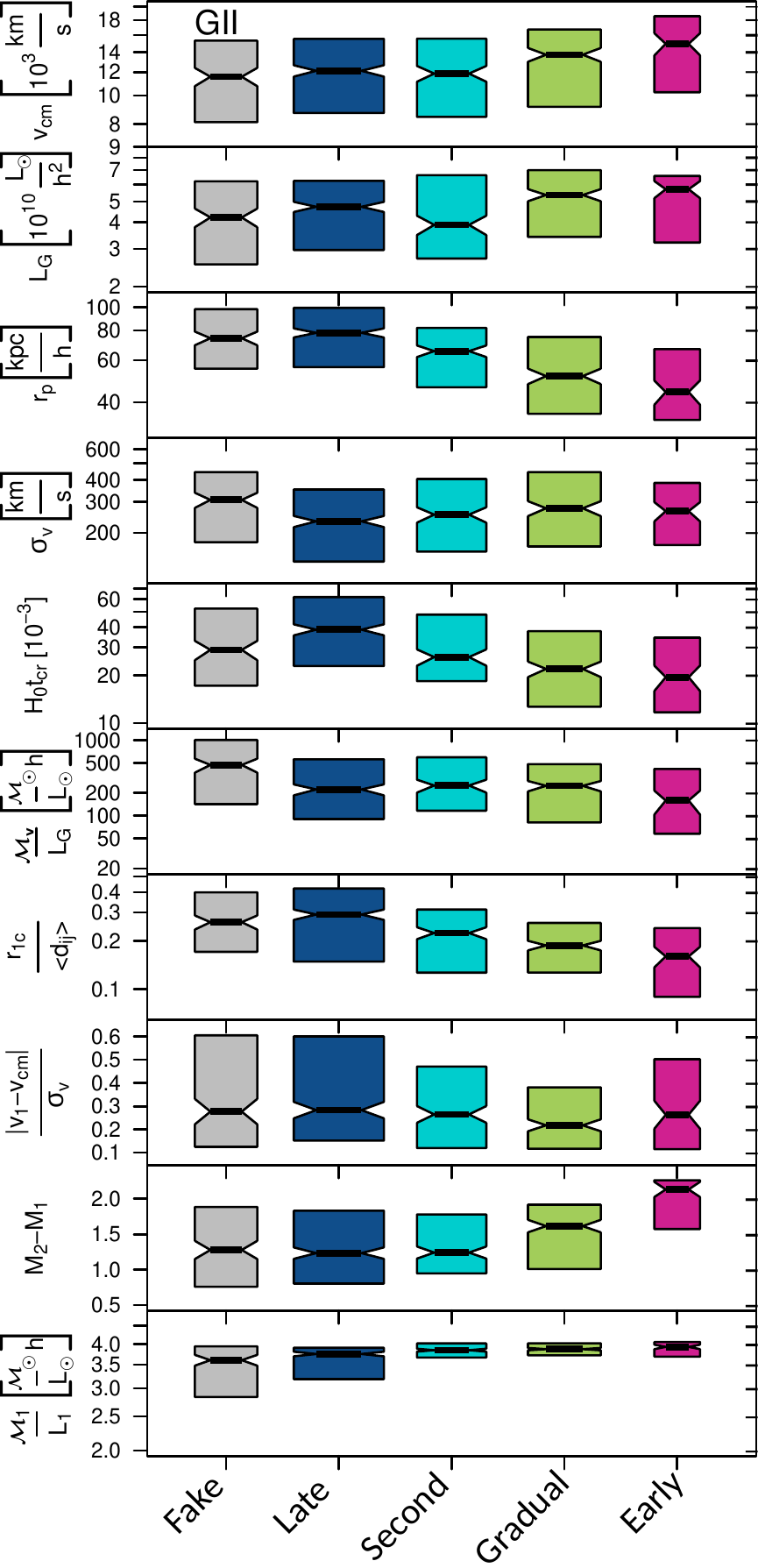}
    \caption{\label{fig:boxgii} 
    Distribution of  the  observable properties of $z$=0  mock 4-galaxy CGs (with no splitting between Real and CAs) as a function of the assembly class for the GII SAM (\emph{second to last columns}). 
    We also include the properties for the sample of CGs previously classified as Fake CAs (\emph{first column}). 
    The properties, from \emph{top} to \emph{bottom}, are 
    distance (median line-of-sight velocity), 
    group luminosity, projected radius of the minimum enclosing circle,
    group line-of-sight velocity dispersion, 
    dimensionless group crossing time, 
    group virial theorem mass-to-light ratio, 
    dimensionless brightest galaxy projected group-centric distance, dimensionless brightest galaxy line-of-sight velocity offset, $r$-band magnitude gap between the two most luminous galaxies, and brightest galaxy stellar mass-to-light ratio.
    The \emph{waists}, tops and bottoms of the \emph{coloured boxes} indicate the median values, and the 75th and 25th percentiles, respectively, the \emph{widths} of the boxes are proportional to the numbers of CGs, while \emph{notches} around the waists show the $95\%$ confidence interval on the medians.
    }
    \end{center}
\end{figure}
Figure~\ref{fig:boxgii} shows the distribution of different properties of CG4s in the GII SAM in form of boxplot diagrams (similar plots for CG4s in the other four SAMs are shown in Figure~\ref{fig:box}). The dimensionless group crossing time and 
    group virial theorem mass-to-light ratio are measured as \cite{DiazGimenez+12}:
    $H_0\,t_{\rm cr} = 50\,(\pi/\sqrt{3})\,(h\,d_{ij}/\sigma_v)$ and 
    $M_{\rm VT}/L = 3\pi\,\left\langle d_{ij}^{-1}\right\rangle^{-1}\,\sigma_v^2/(G\,L)$, where $d_{ij}$ is the median inter-galaxy separation projected on the sky.

Analysing the median values obtained from this figure\footnote{If the notches in two boxplots do not overlap, one can conclude  with $95\%$ confidence that the corresponding  medians are  different \citep{boxplot78,boxplot14}.}, some dependence of these properties on the assembly class is observed. 
CG4s in the {\tt Early} class show the largest magnitude gaps, as well as the lowest projected sizes, crossing times 
and  brightest galaxy relative projected offsets. Those attributes are actually a good reflection of their early formation times. On the other hand, CG4s recently formed show the opposite behaviour, with the {\tt Late} class showing the smallest magnitude gaps, as well as the largest projected sizes, crossing times, and brightest galaxy relative projected  offsets. 

In general, there seems to be a progression in the median properties between the four group assembly channels, following the trend: {\tt Early Assembly}, {\tt Gradual Contraction}, {\tt Late Second Pericentre}, {\tt Late Assembly}. 

We have also included in this figure the properties of those CG4s classified as Fake to have the complete picture of all CG4s in the main sample. In most of the properties, Fake CG4s behave similarly to the recently formed CG4s, except for the crossing times where Fake CG4s have lower values.

While Fig.~\ref{fig:boxgii} shows this analysis only for the GII SAM, roughly the same trends are observed for all the remaining SAMs (see Figure~\ref{fig:box}).
Moreover, Figs.~\ref{fig:boxgii} and \ref{fig:box} show that all SAMs  display weaker but apparently significant trends for {\tt Early} assembly histories for CGs of higher group velocity dispersion and higher brightest galaxy 
stellar mass-to-luminosity ratio. In contrast, contradictory trends among the CG4s built from the different SAMs are found for group luminosity and virial-theorem-mass-to-luminosity ratio.

\section{Summary and Discussion}
\label{sec:discus}
In this work, we studied the different assembly channels that give rise to the compact groups of galaxies (CGs) of galaxies identified at present using an automatic algorithm that follows  Hickson's well-known recipes \citep{Hickson82,Hickson92}.
Following our previous work \citep{paper1}, we identified CGs in several mock lightcones constructed from the Millennium Simulations using different SAMs. Since our study involves following each member of a CG back in time, we simplified our analysis by only selecting
CGs with four members (CG4s), and by only following their most massive progenitors. We thus analysed the assembly histories of over 6000 CG4s in the outputs of five SAMs.

To start with, we performed a preliminary analysis, where we visually inspected a small subset ($\sim $350 CG4s) of CG assembly histories, using the evolution of the 3D physical distances of galaxy members to the group centre of mass. This allowed us to differentiate four different basic assembly channels that we named
 {\tt Early Assembly},
 {\tt Gradual Contraction},
 {\tt Late Assembly},  and
 {\tt Late Second Pericentre}  (Fig.~\ref{fig:evoltest}). The first two channels encompass relatively old systems, while the last two are characterised by rather late assembly. 
 We notice that there is a wide spectrum of possible variations for the evolution of CG4s, and some of them evolve in ways that are at the boundaries between channels. Therefore, our classification is most probably a simplification.
 From the analysis of the test sample, we observed a tendency of CG4s to be predominantly recently formed (Fig.~\ref{fig:test}).
 
We then extended this analysis to the full sample of CG4s, by defining an automatic classification using some of the features measured on the evolution of the physical 3D distance to the group centre of mass of the galaxy with the fewest significant pericentres (see Fig.~\ref{fig:visual}). This automatic classification achieved a 96 per cent average success rate in recovering the visual classification. 

Applying our automatic classification on the full sample of 6000 CGs, we found that recently formed CGs are the dominant assembly class, regardless of the SAMs. 
When analysing CGs classified as Reals according to the classification scheme of \cite{DiazGimenez&Mamon10}, we observed that they have, on average, an important component of older systems, whereas CGs that are Chance Alignments according to the \citeauthor{DiazGimenez&Mamon10} classification are preferentially dominated by recently assembled systems (Fig.~\ref{fig:full}).

One may wonder what is the meaning of assembly classes of CGs that are chance alignments. There is no bimodality in the 3D properties of CGs: as seen in fig.~6 of \cite{DiazGimenez&Mamon10},
our distinction between Reals and CAs is based on a somewhat arbitrary cut in the smooth distribution of mock CGs in the alignment versus size plane. Many CAs barely missed being classified as Reals. Therefore, there is no strong reason to exclude CAs from the analysis.

Among CGs, Reals have greater number density and presumably greater mass density than the CAs. The shorter assembly time for the Reals relative to the CAs is thus related to the shorter assembly times in systems that are denser at $z=0$.

Cosmological parameters play a strong role on CGs.
In Paper I, we had found that the space density of CGs decreases strongly when increasing the density of the Universe. We interpreted this result as the greater contamination of the CG isolation criterion in high-density universes. We had also noted that the fraction of CGs that are CAs increases with $\Omega_{\rm m}$, especially for the CAs that extend in real space beyond the parent group of the CG, which indeed should occur more frequently in high-density universes.
In this work, we find that the density of the Universe plays a major role on CG4 assembly.
Denser flat universes lead to later CG assembly (left panels of Fig.~\ref{fig:earlies}). This is the consequence of greater late infall in denser flat universes \citep{vandenBosch02}.

We also find a marginally strong trend of increasing fraction of {\tt Earlies} in Real CGs extracted from SAMs built from high $\sigma_8$ cosmologies (middle column of panels of Fig.~\ref{fig:earlies}).
Increasing the normalisation of the power spectrum leads to more structures at all scales. In Paper~I, we found that increasing $\sigma_8$ leads to a higher fraction of CGs that are Real. 
Figure~\ref{fig:full} indicates that Real CG4s assemble earlier than CAs. Thus the weak positive correlation between fraction of early assembly with $\sigma_8$ may be the consequence of the higher fraction of Reals and their earlier assembly.

The fraction of early-assembling CGs may depend on the SAM considered.
In comparison with DLB, the G11 SAM based on the same cosmological parameters, leads to a similar fraction of {\tt Earlies} (Fig.~\ref{fig:earlies}).
In comparison with G13, H15 has fewer {\tt Earlies}, but it is not clear whether this difference is caused by the SAM physics or by the different cosmological parameters. In fact, G11 and G13 have mostly the same SAM but different cosmological parameters and they show indeed differences in the fraction of {\tt Earlies}. This suggests that SAM physics only plays a minor role in the CG assembly, for given cosmological parameters. 

The resolution of the simulation appears to play a role. We compare the fractions of {\tt Earlies} between the high-resolution GII SAM and the comparable lower-resolution G11 SAM (same study and cosmological parameters). We find that the fraction of {\tt Earlies} among all CGs is one-third (and significantly) higher for the high-resolution GII. 
This is expected, because in higher-resolution simulations small or low-mass structures form earlier (as do galaxies, see, e.g., fig.~11 of \citealt{Tweed+18}).

Another relevant issue in the SAMs is the way they deal with the orphan galaxies, whose positions are extrapolated  with analytical recipes. Thus, if a key galaxy is   
an orphan,
then the assembly channel for its group could be based on such an imprecise extrapolation.
Figure~\ref{fig:orph} indicates
that $\sim 30 \%$ of the CG4 assembly channel classifications were based on orphan key galaxies. In such cases, galaxies in {\tt Second, Gradual } and {\tt Early} CGs became orphans once the systems were already small. Key galaxies in {\tt Late} CGs  were located farther from the group centre when they became orphans, but the percentage of {\tt Late} CGs with orphan key galaxies is rather low. Therefore, we find that having an orphan galaxy as key galaxy to classify the assembly of CGs does not introduce biases in our results.

Finally, we also investigated if the formation channel of CGs has an effect on their observational properties. From the analysis of several observational properties, there seems to be a progression in the median properties: earlier formed CGs tend to be the smallest, densest (shortest crossing times), with the largest magnitude gap between the two brightest members, while  more recently formed CGs show the opposite behaviour (Figs.~\ref{fig:boxgii} and \ref{fig:box}).

There are similarities between the correlation of assembly times and observational properties for CG4s and for groups in general.
Since the pioneering work of \cite{jones03}, old galaxy groups have been directly associated with systems showing a large magnitude gap between the two brightest galaxies. This association has been confirmed using galaxies in mock catalogues built from SAMs, not only using the gap between the two brightest galaxies (e.g., \citealt{Dariush+07}; \citealt*{DiazGimenez+08}), but also using the gap between the first and fourth brightest galaxies (e.g., \citealt{Dariush+10}; \citealt*{Kanagusuku+16}). 
Other physical quantities have been found to correlate with the epoch of group assembly.
\cite*{Khosroshahi+07} found that older galaxy systems (with large magnitude gap) have a more concentrated halo compared with those recently formed.
In the same vein, \cite{ragone10} found that more concentrated groups form earlier in cosmological hydrodynamical simulations.
In our work, we found that the four galaxy members in early formed CGs are clearly confined to a smaller region of space than those in recently formed CGs.  

\cite{Raouf+14} argued that group assembly history is best measured with a combination of observational parameters: they considered the group and brightest group galaxy (BGG) luminosities and  the physical offset of the BGG  from the luminosity centroid.
\cite{Farhang+17} found that CGs acquire half their mass earlier than normal groups, but later than  Fossil groups. 
Our results have shown that the findings valid for early-forming galaxy groups also hold for early-assembly  CGs (at least for median values), which are
denser (shorter crossing times) and whose 
BGGs have greater magnitude gaps, and lie closer to the luminosity centroid, than later-assembly CGs.

However, our analysis is different from all these studies, because we do not measure the epoch when half of the CG mass is assembled. Instead, we measure the spatial assembly history, which we split into four formation channels. We found that these   channels represent themselves an evolutionary sequence that are correlated with observational properties in accordance to what was found for the mass assembly history by the studies mentioned above.

Unfortunately, the median observational properties of CGs split into each of the four formation channels are not sufficiently different to reliably infer the formation channel of individual CGs.

Some issues remain to be clarified. What prevents {\tt Early Assembly} CGs to gradually shrink in size by dynamical friction and inelastic galaxy-galaxy encounters? Will our results differ when we analyze mock catalogues built from cosmological hydrodynamical simulations (as rapidly done by \citealt{Hartsuiker&Ploeckinger20}), 
which now have sufficient spatial resolution to achieve better small-scale physics compared to SAMs?  
Do the formation channels of 4-member CGs differ from those of other (less dense) groups of 4 members? 

\section*{Acknowledgements}
{We thank the anonymous referee for useful comments, as well as Matthieu Tricottet.
We also thank the authors of the SAMs for making their models publicly available. 
The Millennium Simulation databases used in this paper and the web application providing online access to them were constructed as part of the activities of the German Astrophysical Virtual Observatory (GAVO).  
This work has been partially supported by Consejo Nacional de Investigaciones Cient\'\i ficas y T\'ecnicas de la Rep\'ublica Argentina (CONICET) and the Secretar\'\i a de Ciencia y Tecnolog\'\i a de la Universidad de C\'ordoba (SeCyT).}

\section*{Data Availability}
The data underlying this article were accessed from \url{http://gavo.mpa-garching.mpg.de/Millennium/}. Galaxy lightcones built in Paper~I and used in this work are accessed from \url{https://doi.org/10.7910/DVN/WGOPCO} \citep{data_paperI}.
The derived data generated in this research will be shared on reasonable request to the corresponding authors.

\bibliography{refs}

\appendix

\section{Compact group virialization redshift if they remain as dense until present} 
\label{sec:zvir}
We estimate here the redshift $z_{\rm vir}$ when CGs virialize if they remain at
the same physical density until present. 
The mean density of the CG is 
\begin{equation}
\overline{\rho} = \Delta_{\rm CG} \,\rho_{\rm crit}(0) = \Delta(z_{\rm vir})
\rho_{\rm crit}(z_{\rm vir}) \ ,
\end{equation}
where $\rho_{\rm crit}(z) = 3 H^2(z)/(8\pi G)$ is the critical density.
The CG virialization redshift is therefore the solution of
\begin{equation}
\Delta(z) \,E^2(z) = \Delta_{\rm CG} \ ,
\label{eqzvir}
\end{equation}
where $E(z) = H(z)/H_0 = \sqrt{\Omega_{\rm m}(1+z)^3 + 1 - \Omega_{\rm
    m}}$ for a flat Universe with current density parameter $\Omega_{\rm m}$.
According to the values listed in table~5 of \cite{DiazGimenez+12},
the overdensity of CGs is $\Delta_{\rm CG} = 3\,({\cal M}_{\rm VT}/L_R)\,L_R /
(4\,\pi\,r_p^3) = 24\,000$ and 28\,000, for the
median CGs in the 2MASS catalogue \citep{DiazGimenez+12} and mock CGs (from the G11 SAM) respectively (all considered after velocity
filtering). Here, we measured the CG 
density within the sphere of radius 
equal to the smallest circumscribed radius (this assumes that the typical CG
is spherical).
Solving equation~(\ref{eqzvir}), we deduce $z_{\rm vir}=7.1$ (2MCG)
and 7.6 (mock CGs).
However, CGs may be more extended along the line-of-sight than viewed in projection. 
Adopting a CG radius of $2\,r_{\rm p}$ instead of $r_{\rm p}$, the
overdensities are 8 times lower, and the virialization redshifts are now
$z_{\rm vir}=3.0$
and 3.2 respectively.

\section{Analysis of CGs in all semi-analytical models}
\subsection{Assembly histories of CA CGs}
\label{nnfull}

\begin{table}
\tabcolsep 2.5pt
\centering
\caption{Number of $z$=0 CAs and percentages within each CA subclass}
\begin{tabular}{lrccccc}
    \hline
        SAM & & \multicolumn{5}{c}{CAs}  \\
         \cline{3-7}
         && \multicolumn{1}{c}{\#} & \%\,Fake & \%\,3+1 & \%\,2+2 & \%\,2+1+1  \\
        \hline
All & & 4423 & $27\pm 1$ & $52\pm 1$ & $8\pm 1$ & $13\pm 1$ \\
\hline
DLB & & 1320 & $28\pm 2$ & $52\pm 3$ & $7\pm 1$ & $13\pm 2$ \\
G11 & & 1040 & $24\pm 3$ & $54\pm 3$ & $7\pm 2$ & $15\pm 2$ \\
G13 & & 710 & $28\pm 3$ & $48\pm 4$ & $9\pm 2$ & $15\pm 3$ \\
H15 & & 579 & $30\pm 4$ & $47\pm 4$ & $8\pm 2$ & $15\pm 3$ \\
GII & & 774 & $25\pm 3$ & $56\pm 3$ & $9\pm 2$ & $10\pm 2$ \\
    \hline
    \hline
\end{tabular}
\parbox{8.4cm}{Note: Each percentage $p$ is quoted as $p\pm ci$ where $ci$ is the 95\% binomial confidence interval computed as $\pm (1.96\times \sqrt{f(1-f)/N_{\rm CG}})\times 100$ where $f=p/100$.
}
\label{tab:subcas}
\end{table}

For a more detailed analysis, we defined CA sub-classes as follows:  
\begin{itemize}
    \item {\bf 3+1:} $r_{\rm max} < D_{\ast}$ and $r_{\rm min} < d_{\ast}$ and $r_{\rm max3} \le d_{\ast}$,
    \item {\bf 2+2:} $r_{\rm max} < D_{\ast}$ and $r_{\rm min} < d_{\ast}$ and $r_{\rm max3} > d_{\ast}$ and $r_{ij} \le d_{\ast}$,
    \item {\bf 2+1+1:} $r_{\rm max} < D_{\ast}$ and $r_{\rm min} < d_{\ast}$ and $r_{\rm max3} > d_{\ast}$ and $r_{ij} > d_{\ast}$,
    \item {\bf Fake:} $r_{\rm max} \geq D_{\ast}$ or $r_{\rm min} \geq d_{\ast}$,
\end{itemize}
where $r_{\rm max}$ and $r_{\rm min}$ are the maximum and minimum inter-particle separation using the 4 members of the group, respectively, $r_{\rm max3}$ is the maximum inter-particle separation between the 3 closest members of the group, $r_{ij}$ is the inter-particle separation between the 2 members that do not define the pair with $r_{\rm min}$, and the thresholds are $D_{\ast}= 1 \ h^{-1} \, \rm Mpc$  and $d_{\ast}=200 \ h^{-1} \, \rm kpc$.
Table~\ref{tab:subcas} displays the percentages of the different CA sub-classes for each SAM. 

\begin{figure}
    \centering
    \includegraphics[width=\hsize,trim=0 0 20 30]{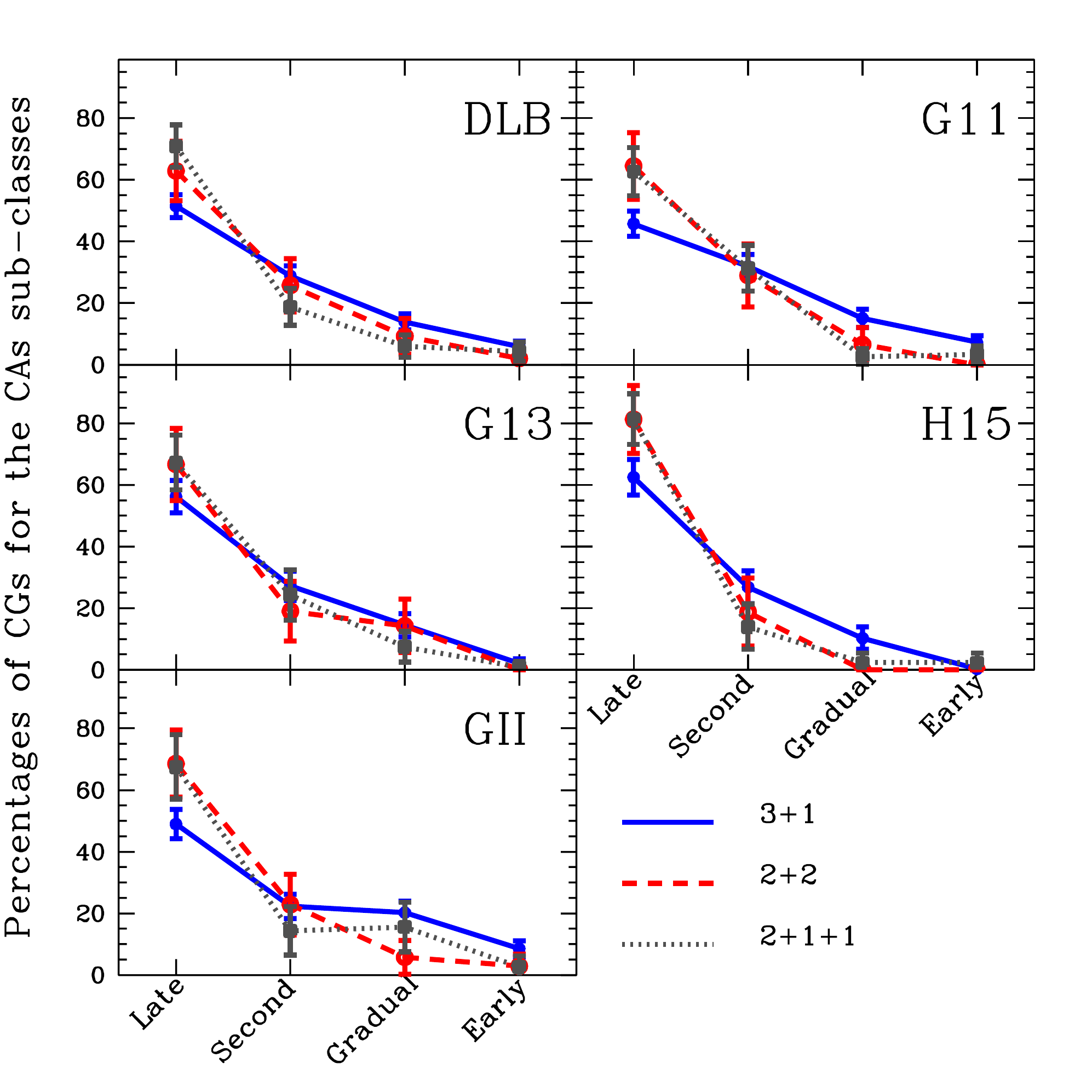}
    \caption{Same as Fig.~\ref{fig:full}, but only for the 4-galaxy CGs that are chance alignments, split by sub-class.
    }
    \label{fig:cassub}
\end{figure}

We checked how the assembly histories differ among the different sub-classes of CAs (excluding Fakes).
Figure~\ref{fig:cassub} shows the percentages of CG4s for the CA sub-classes as a function of assembly class. All the CA sub-classes roughly follow the general trend previously observed for the full sample of CAs,
with the main contribution to {\tt Gradual} and {\tt Early} classes of the CAs coming from the dominant ``3+1'' sub-class. The sub-classes with one or two galaxy pairs are mainly recently formed in all SAMs.
In summary, the richer the subsystem, the earlier it tends to assemble, from pairs (2+2 or 2+1+1 CAs) to triplets (3+1 CAs) to quartets (Reals).

Interestingly, the key galaxy of a CA, which by definition is the galaxy with the fewest orbits, is not necessarily an outlier at $z=0$ (for 3+1 CAs) or one of them (for 2+2 or 2+1+1 CAs).
So the classification of CA assembly histories, which is a function of the orbit of the key galaxy, is not necessarily tied to an outlier but to one of the galaxies of the physical subsystem (i.e triplet in a 3+1 CA). However, the key galaxy of a CA is more likely to be an outlier galaxy, since to first order, galaxies further out live in lower mean densities, hence have the longest orbital times and the fewest number of orbits. Nevertheless, the fraction of CAs classified as {\tt Late} or {\tt Second} is less than two-thirds.
On the other hand, finding that the 3+1 CAs are the main contributor to Earlies is expected, because 
3+1 CAs have only one outlier and are thus less likely to have their key galaxy be an outlier than 2+2 or 2+1+1 CAs.

\subsection{Observational properties versus SAM}
\begin{figure*}
\begin{center}
\includegraphics[width=23cm,angle=-90]{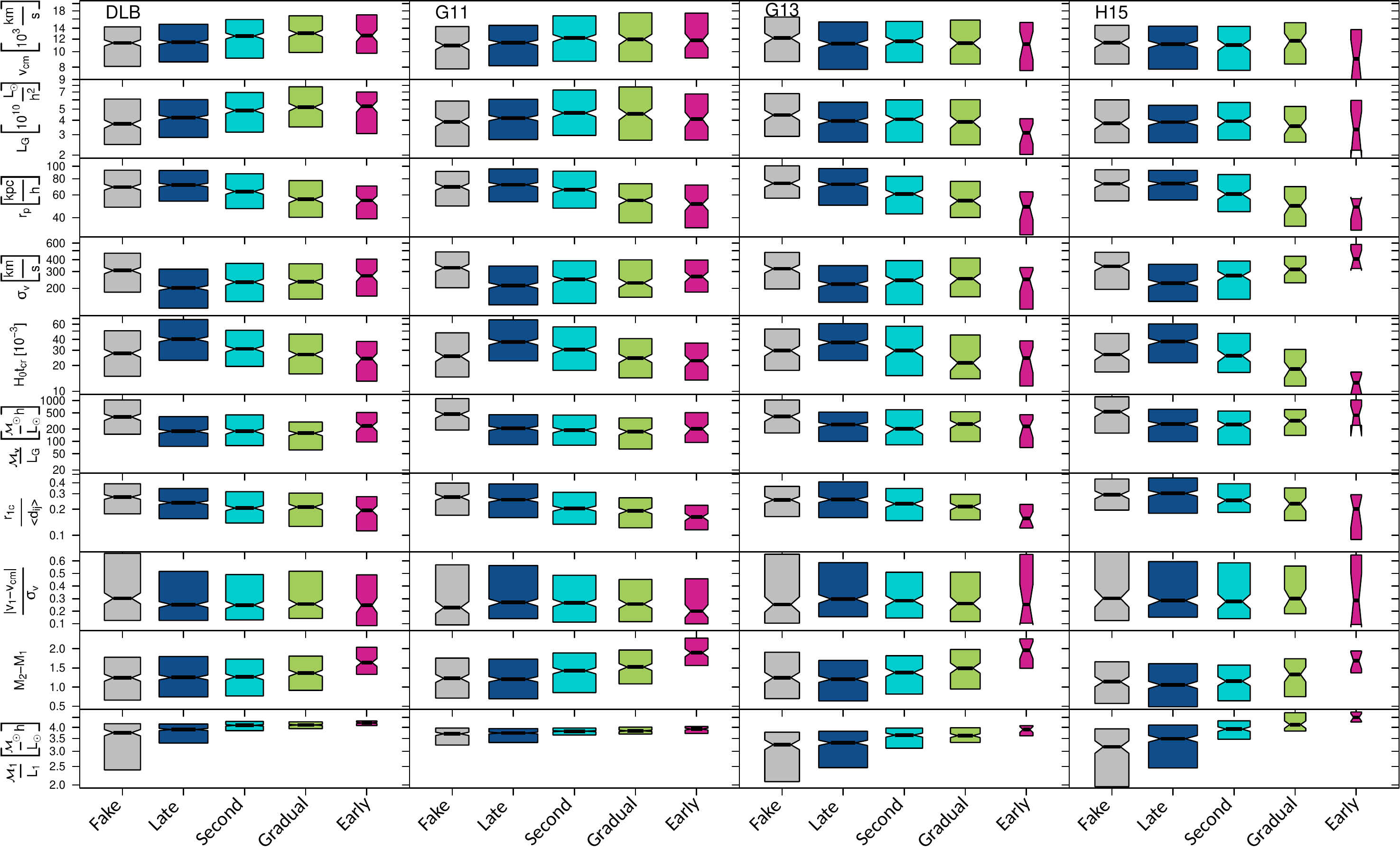}
\caption{\label{fig:box} 
Same as Figure~\ref{fig:boxgii}, but for the CGs built from the four other semi-analytical models.}
\end{center}
\end{figure*}

Figure~\ref{fig:box} shows the properties of CGs in each assembly class for the SAMs of DLB, G11, G13 and H15.

\end{document}